\documentclass{aastex631}
\usepackage{natbib}
\usepackage{xspace}
\usepackage{tabularx}
\usepackage{multirow,makecell}

\newcommand{\logt}{$\log T$\xspace}
\newcommand{\emps}{EM$_{\rm{PS}}$\xspace}
\newcommand{\mt}{$\langle T\rangle$\xspace}
\newcommand{\mtps}{$\langle T\rangle_{\rm{PS}}$\xspace}

\shorttitle{Current Sheet Plasmas}
\shortauthors{Gou \& Reeves}

\accepted{June 27, 2024}
\submitjournal{ApJ}

\begin{document}

\title{Thermal Properties of Current Sheet Plasmas in Solar Flares}

\correspondingauthor{Tingyu Gou}
\email{tingyu.gou@cfa.harvard.edu}

\author[0000-0003-0510-3175]{Tingyu Gou}
\author[0000-0002-6903-6832]{Katharine K. Reeves}
\affiliation{Center for Astrophysics $|$ Harvard $\&$ Smithsonian, 60 Garden Street, Cambridge, MA 02138, USA}

\begin{abstract}
The current sheet is an essential feature in solar flares and is the primary site for magnetic reconnetion and energy release. Imaging observations feature a long linear structure above the candle-flame-shaped flare loops, which resembles the standard flare model with the current sheet viewed edge-on. We investigate the thermal properties of plasmas surrounding the linear sheet during flares, using EUV observations from the Atmospheric Imaging Assembly (AIA) onboard the Solar Dynamics Observatory (SDO). The differential emission measure (DEM) analyses show evidence of high temperatures in the plasma sheets (PSs), containing hot emissions from only a narrow temperature range, suggestive of an isothermal feature. The sheet's temperature remains constant at different heights above the flare arcade, peaking at around \logt=7.0--7.1; while the well-studied 2017 September 10 X8.2 flare exhibits as an exception in that the temperature decreases with an increasing height and peaks higher (\logt=7.25) during the gradual phase. Most PS cases also hold similar emission measures and thicknesses; while the PS's emissions drop exponentially above the flare arcade, the sheet thicknesses show no significant height association as for all the measurements. The characteristics of isothermal and steady temperature suggests balanced heating and cooling processes along the current sheet, particularly additional heating may exist to compensate for the conductive and radiative cooling away from the reconenction site. Our results suggest a steady and uniform sheet structure in the macroscopic scale that results from flare reconnection.
\end{abstract}


\section{Introduction} \label{sec:intro}

Solar flares are explosive and energetic phenomena in the solar atmosphere, which can release a huge amount of magnetic energy within a short time. The energy release process is usually attributed to magnetic reconnection that occurs at the current sheet in the wake of the eruption \citep[e.g., see reviews by][]{Priest2002,Shibata2011,Benz2017}. The classic flare model in two dimensions \citep{Carmichael1964,Sturrock1968,Hirayama1974,Kopp1976,LinJ2000} features a linear structure connecting the bottom of an erupting magnetic flux rope and the tip of candle-flame-shaped flare loops underneath, which shows an eruptive picture with a current sheet viewed from an edge-on perspective \citep{LinJ2015}. During the eruption, magnetic inflows are continuously brought into the current sheet, where magnetic reconnection occurs and produces bi-directional outflows moving upward or downward along the sheet \citep[e.g.,][]{Forbes1996,LinJ2000}. Meanwhile, magnetic free energy stored in the coronal magnetic field is rapidly released and converted into thermal and kinetic energies that are used for plasma heating and particle acceleration in the flare. Thus, the current sheet is an essential feature to power flares, and its properties are important to understand the magnetic reconnection and associated energy release processes.

In remote-sensing observations of solar flares, currently it is not possible to observe the electric currents in the corona, but the plasmas surrounding the current layer can give a sense of the current structure itself thus are directly relevant to the reconnection process. During eruptive flares, imaging observations sometimes feature a thin, linear structure in white light, EUV, or X-ray passbands \citep[e.g.,][]{Ko2003,Ciaravella2008,Liu2010,Seaton2018,Gou2019}, inside which upward or downward moving outflows are also detected \citep{Savage2010,LiuR2013,Zhu2016,Longcope2018,Gou2020,Yu2020}. These observations resemble the standard flare model and provide clear indications for the existence of a reconnection current sheet viewed edge-on. When an eruption is viewed face-on, the current sheet region exhibits as a broad fan of diffusive plasmas above the post-flare arcade, where dark voided structures known as supra-arcade downflows are observed to intermittently move toward the flare arcade \citep{McKenzie1999,Innes2003,Savage2012,Reeves2017,Xue2020,Awasthi2022}. These observations provide a complementary insight into the flare current sheet. Thermal diagnostics from multi-wavelength observations reveal that the plasmas surrounding flare current sheet, exhibiting as either a long thin sheet or a supra-arcade fan, are associated with high temperatures \citep[e.g.,][]{Gou2015,Reeves2017,Seaton2017}. These structures are mostly visible in hot EUV channels such as the 131~\AA\ (primarily from Fe XXI emission line, \logt=7.05) and/or the 193~\AA\ (with contributions from Fe XXIV, \logt=7.25) channels of the Atmospheric Imaging Assembly \citep[AIA;][]{Lemen2012} onboard the Solar Dynamics Observatory \citep[SDO;][]{Pesnell2012}. Spectroscopic observations on the current sheet region show excess broadening in hot spectral lines associated with non-thermal velocities \citep{Ciaravella2008,Warren2018,LiY2018,French2020,Shen2023}. Although observations of current sheet cases are still rare, these results suggest the existence of heating and turbulent processes within the current sheet, which are immediately driven by magnetic reconnection. 

Magnetic reconnection in flares can occur in a bursty or fractal fashion with highly time-dependent characteristics. Theoretical and numerical studies demonstrate that the current sheet during fast reconnection can be highly fragmental and turbulent, with continuous formation and injections of small magnetic islands, also termed plasmoids \citep{Loureiro2007,Ni2015,Dong2022,Ye2023,Wang2023}. These plasmoids may play an important role in modulating the rate of magnetic reconnection and energy transfer as multiple reconnection sites are generated during the thinning of current sheet \citep{Shibata2001,Dong2022}. In observations, efforts are made to understand the fast reconnection process in solar flares. Studies using high-resolution imaging from SDO/AIA reveal sub-structures of a flaring plasma sheet where a linear structure breaks up into multiple plasmoids \citep[e.g.,][]{Takasao2012,Gou2019}, and the ejection and coalescence of plasmoids in various scales suggest a fractal fashion of the current sheet \citep{Gou2019}. However, direct imaging of plasmoids in flaring current sheets are very rare, much fewer than the cases of current sheet candidates in EUV, although the tearing mode instability is expected to occur as the critical value of length-to-thickness ratio has been reached \citep{Furth1963}. On the other hand, the apparent thickness of current sheet in observations is much greater than that in theoretical and numerical studies \citep{LinJ2015}, which require a microscopic plasma scale to produce the anomalous resistivity. These different features between models and observations appeal for more detailed investigations on the flare current sheet.

In this study, we focus on the long, thin plasma sheet structures observed in EUV during eruptive solar flares. High-resolution imaging by SDO/AIA features a linear plasma sheet (PS) above the flare loops in some flare events, especially for those occurring near the solar limb. Simultaneous multi-wavelength observations of AIA allow us to study the thermal properties of plasmas surrounding the flare current sheet. A small number of individual case studies have been reported before, and most of them focus on the extremely long and dense PS feature observed during the 2017 September 10 X8.2 flare owing to its favourable viewing angle and rich observational data from various instruments. In this work, we search and investigate PSs observed in a collection of several flares which show prominent bright emission features. Our results show that most PS features exhibit similar thermal properties while the 2017 flare is an exception. Our analyses suggest an isothermal and uniform sheet structure at macroscopic scales in spite of its possible turbulent and fragmented characteristics at the micro level. We present the method and detailed analyses in Sections~\ref{sec:methods} \& \ref{sec:analyses} and discuss the results in Section~\ref{sec:discussion}.

\section{Events and Methods} \label{sec:methods}

\begin{figure}[htbp]
	\centering
	\includegraphics[width=\textwidth]{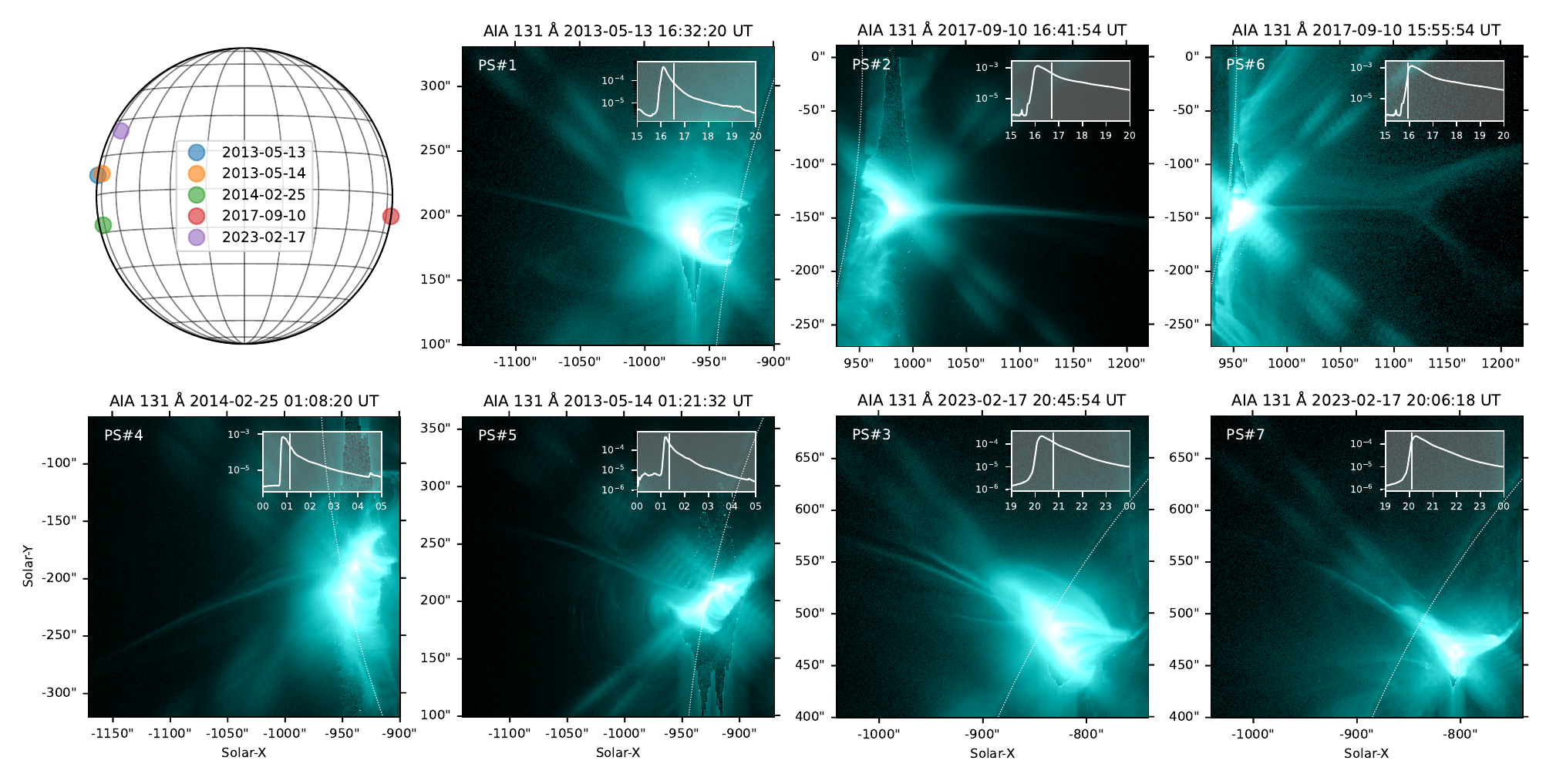}
	\caption{\small SDO/AIA observation of the solar flares and PS structures. The locations of flares on the Sun are indicated by colored circles in the top-left panel. The AIA 131~\AA\ snapshots are composite images constructed from the combination of a pair of long and short exposure time data. The dotted white curves in AIA images indicate the solar limb. The inserted plots in each map show the corresponding GOES 1--8~\AA\ SXR flux of the flare with a duration of 5 hours, where the vertical solid line marks the time of the AIA 131~\AA\ map when the observed PS structure is selected for study.}
	\label{fig:flares}
\end{figure}

\begin{table}[htb]
	\centering
	\caption{List of flares and PS cases.} \label{tab:flares}
	\begin{tabular}{ccccccc} 
		\hline\hline
		Flares & GOES Class & Start\footnote{All times are in UT.} & Peak & PS Cases & Location (AR) & CME Speed\footnote{The CME speeds are linearly fitted results (in km/s) obtained from the 
        \href{https://cdaw.gsfc.nasa.gov/CME_list/}{CDAW SOHO/LASCO CME Catalog}.} \\ 
        \hline\hline
		2013 May 13 & X2.8 & 15:48 & 16:05 & \#1~~16:32 & N08E89 (11748) & 1850 \\ [1ex] \hline
        2017 Sep 10 & X8.2 & 15:35 & 16:06 & \makecell[{}{>{\parindent-1.8em}m{1.5cm}}]{\#6~~15:55 \\ \#2~~16:41} & S08W88 (12673) & 3163 \\ \hline
        2023 Feb 17 & X2.3 & 19:38 & 20:16 & \makecell[{{>{\parindent-1.8em}m{1.5cm}}}]{\#7~~20:06 \\ \#3~~20:45} & N25E67 (13229) & 1315 \\ \hline
		2014 Feb 25 & X4.9 & 00:39 & 00:49 & \#4~~01:08 & S12E77 (11990) & 2147 \\ [1ex] \hline
		2013 May 14 & X3.2 & 00:00 & 01:11 & \#5~~01:21 & N08E77 (11748) & 2625 \\ [1ex]
        \hline\hline
	\end{tabular}
\end{table}

We study the properties of flare PS structures using EUV observations from SDO/AIA. Favorable flare events are selected that show a long, linear feature above post-flare loops in hot AIA passbands such as 131~\AA. Such events require an almost edge-on perspective of the AIA instrument where the line of sight is largely parallel to the axis of the post-flare arcade. Favorable observations showing long PS structures are limited. Here we investigate sheet structures nicely observed in five solar flares, which are shown in Figure~\ref{fig:flares} and listed in Table~\ref{tab:flares}. All of these events are intense X-class, long-duration flares occurring near the solar limb, and they all are eruptive events associated with fast CMEs.

During the flares under study, the long PS features are most evident shortly after the soft X-ray (SXR) flux peaks, i.e., during the early decay phase of the long-duration flares (PSs \#1--5; Figure~\ref{fig:flares}, Table~\ref{tab:flares}). The linear structure is sometimes also visible during the flare impulsive phase, which connects the upper tip of cusp-shaped flare loops and the bottom of an erupting flux rope. Among two of the five flares, we select two PS cases during the impulsive phase, which are clearly identifiable in AIA 131~\AA\ images and have little overlap with either leg of the erupting flux rope (PSs \#6 \& 7; Figure~\ref{fig:flares}). We also carefully check the EUV images to avoid any overlap between the PSs and the diffraction patterns from the bright flare loops off the mesh of the AIA instrument, the latter of which is always inevitable in such intense flares. The seven PS cases under study are shown in Figure~\ref{fig:flares}. SDO/AIA provides high-spatial resolution (with a pixel size of 0.6$\arcsec$, $\sim$0.4~Mm) and simultaneous multi-wavelength EUV observations, which makes it possible to study the thermal properties of plasmas surrounding the flare current sheets.

We adopt the differential emission measure (DEM) method to diagnose the temperature structure of flare plasmas, using data from six AIA EUV channels, i.e., 131~\AA, 94~\AA, 335~\AA, 193~\AA, 211~\AA, and 171~\AA. To reduce the diffraction pattern of AIA telescopes and effects of other stray lights, the AIA data are further processed to level 1.6 by deconvolving images with the instrument point spread function (PSF) before the DEM calculation (see, e.g., Figure~\ref{fig:ps13-aia}). We apply the modified sparse inversion DEM code \citep{Cheung2015,SuY2018} which effectively constrains hot flare emissions, and calculate emission measures (EMs) in the temperature range of \logt= 5.5--7.6 with an interval of $\Delta\log T$ = 0.05. We generate a temperature map by deriving the EM-weighted mean temperature \mt over the whole temperature range (\logt= 5.5--7.6): 
\begin{equation}
	\langle T\rangle=\frac{\sum_{i} \mathrm{EM}(T_i) \times T_i}{\sum_{i} \mathrm{EM}(T_i)}.
\end{equation}
The DEM distribution of optically thin plasmas during a flare usually contains more than one components, where the cooler one is mostly contributed from the foreground and background and the hotter one is from the flaring plasma itself \citep{Hannah2012,Gou2015}. To reduce the effects of background contributions, we define a mean temperature of the PS structure, \mtps, by only focusing on the hot component ($T=a$--$b$) of each PS case \citep[see also][]{Gou2015}:
\begin{equation}
    \langle T\rangle_{\rm{PS}}=\frac{\sum_{i=a}^{b} \mathrm{EM}(T_i) \times T_i}{\sum_{i=a}^{b} \mathrm{EM}(T_i)}.
\end{equation}
Similarly, we calculate the clean emissions from the PS, \emps, by summing up EMs from only the hot component:  
\begin{equation}
	\mathrm{EM}_{\mathrm{PS}}=\sum_{i=a}^{b} \mathrm{EM}(T_i).
\end{equation}
The specific temperature range of the hot DEM component ($T=a$--$b$) for each PS case differs, and they are determined in Section~\ref{sec:analyses} accordingly.

The emissions from the PS structure decrease greatly with an increasing height above the flare arcade. To investigate the height distribution, we use exponential functions to fit both the AIA intensity and \emps profiles, based on an exponential distribution of the density $n_e$ with height $h$ in the solar corona,
\begin{equation}
    n_e (h, T) = n_{e0}~\mathrm{exp}\left[-\frac{h}{H(T)}\right],
\end{equation}
where $H(T)$ is the density scale height. While the observed emissions sum up all density contributions along the line of sight $l$ if assuming a fully ionized plasma in the corona,
\begin{equation}
    \mathrm{EM}(T) = \int n_e^2(T,l)dl,
\end{equation}
we obtain an exponential distribution of the observed AIA intensity ($I_\mathrm{AIA}$) and EM in the form of
\begin{equation}
    I_\mathrm{AIA}~\sim~\mathrm{EM}~\sim~\mathrm{exp} \left [-\frac{2h}{H} \right ].
\end{equation}
We investigate the height distribution of PS emissions in terms of the EM scale height $H$.

The flare loop-top region is usually associated with very dense plasmas around the flare peak, where the DEM inversion would fail to obtain reasonable solutions due to its application for optically thin corona. In addition, to better resolve the long PS structures, we use AIA 131~\AA\ data observed with long exposures, which are often saturated in flare loops with diffraction patterns (e.g., Figure~\ref{fig:ps13-aia}b). The DEM results from these regions should also be ignored. Here in our study, we only focus on the PS region above the flare loop-top. To better visualize, we rotate the AIA maps to place the PS structures either horizontally or vertically in our analyses in the following sections.

\section{Analyses and Results} \label{sec:analyses}

\subsection{2013 May 13 X2.8 flare (PS\#1)}

\begin{figure}[htbp]
	\centering
	\includegraphics[width=\textwidth]{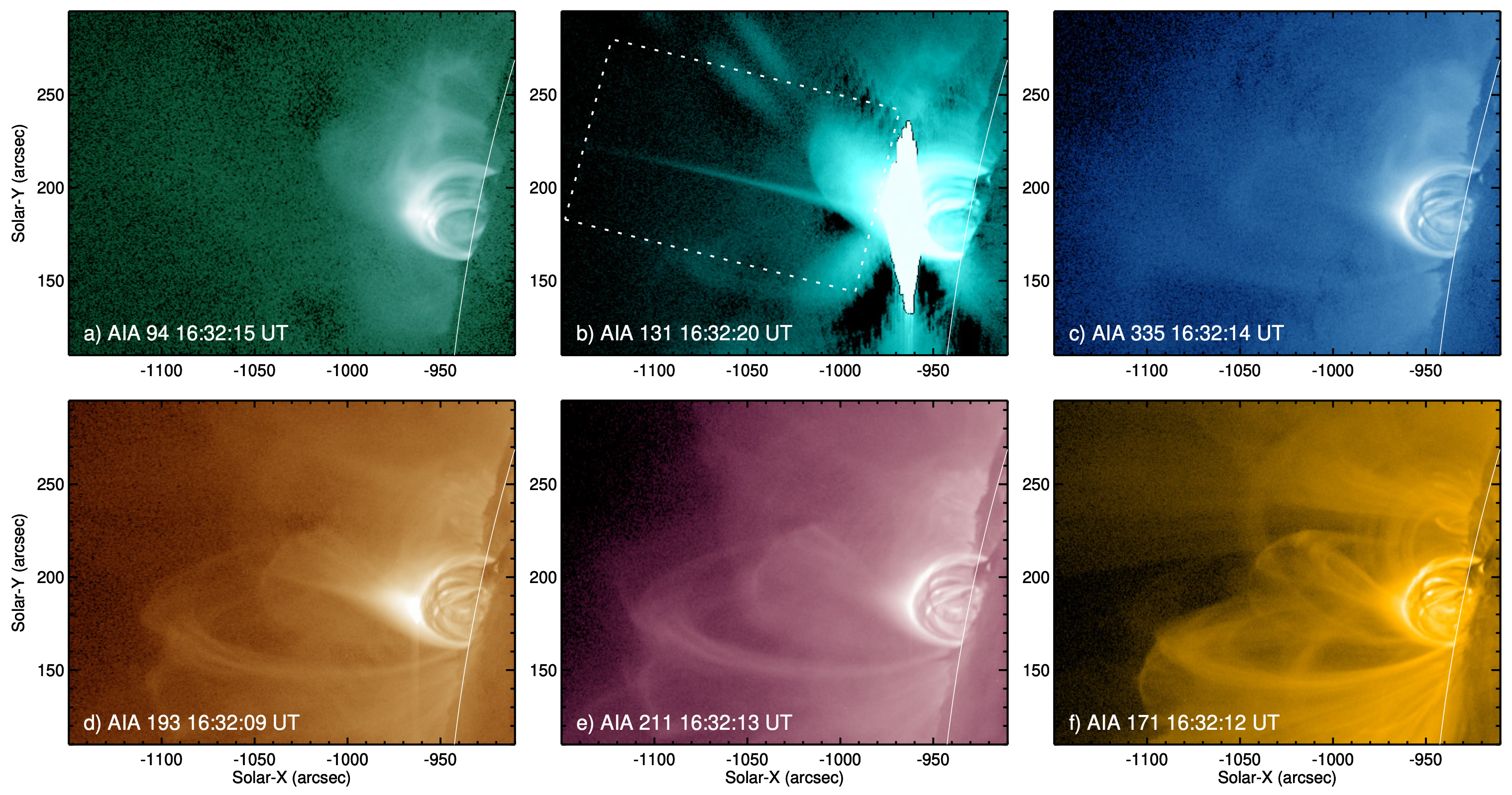}
	\caption{\small SDO/AIA multi-wavelength observations of the 2013 May 13 flare. The AIA images are deconvolved with the corresponding PSF of each wavelength channel. The white curves indicate the solar limb. The dotted rectangular in panel (b) indicates the FOV of maps in Figures~\ref{fig:ps13-dem},\ref{fig:ps13-len-131},\ref{fig:ps13-thi-131}.}
	\label{fig:ps13-aia}
\end{figure}

The 2013 May 13 X2.8 flare occurs in NOAA AR 11748 at the northeast solar limb. The event starts at 15:48~UT and peaks at 16:05~UT, followed by a long-duration ($>$4~hrs) gradual phase. The impulsive phase of the flare is associated with the eruption of a magnetic flux rope, which is observed in AIA 131~\AA\ as a hollow, elliptical feature with its bottom connecting to the cusp-shaped flare loops underneath by a linear structure \citep[see, e.g.,][]{Gou2019}. The eruptive picture is in good agreement with the standard flare model, and the AIA observation provides a nice edge-on view of the limb flare. Shortly after the flare SXR peak, a thin PS appears above the post-flare loops. The PS feature is observable in AIA 131~\AA\ for more than two hours, where a number of supra-arcade downflowing loops are detected as a signature of ongoing magnetic reconnection during the long-duration flare gradual phase \citep[see Figure 5 and the associated movie in][]{Gou2020}.

\subsubsection{Temperature Range}

\begin{figure}[htbp]
	\centering
	\includegraphics[width=\textwidth]{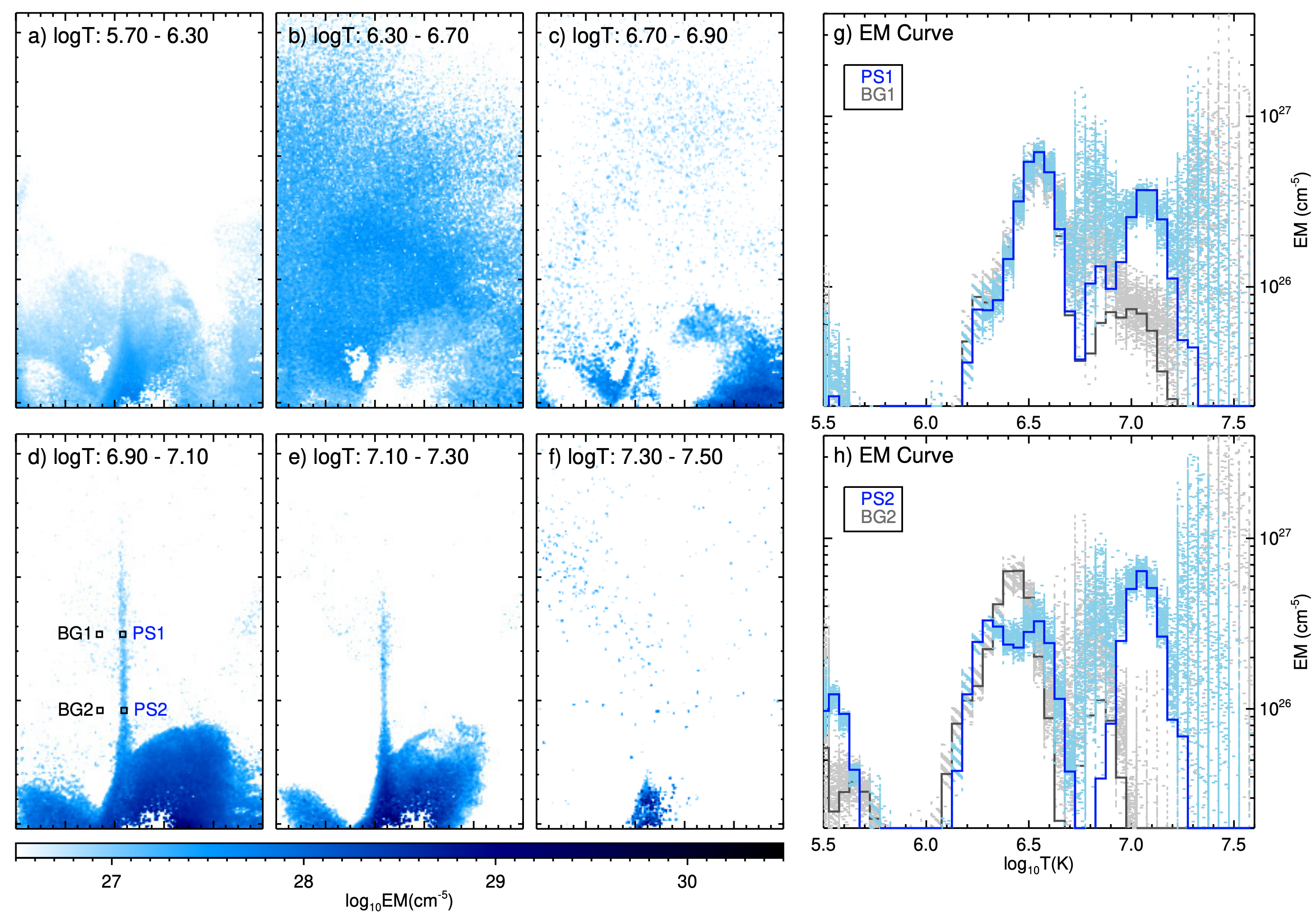}
	\caption{\small DEM results of the 2013 May 13 flare. a--f) EM maps in integrated temperature ranges. Maps are rotated for a better visibility. g,h) Comparison of DEM distributions from the PS and nearby reference locations. The EM curves are sampled from locations indicated by small boxes (averaged over $3\times3$ pixel$^2$) in panel (d), PSs in blue and BGs in black, respectively. The light blue and gray curves are corresponding uncertainties from 250 Monte Carlo simulations.}
	\label{fig:ps13-dem}
\end{figure}

We study properties of the PS at 16:32~UT when it is most evident in AIA (Figure~\ref{fig:ps13-aia}). The long thin feature is most visible in AIA 131~\AA\ and partially in 193~\AA, indicative of hot properties. This characteristic is evidenced by the DEM results in Figure~\ref{fig:ps13-dem}, where the linear PS feature is absent in EM maps with temperatures $\log T<$6.9 and only visible in \logt =6.9--7.3, while only the cusp tip of flare loops appears in temperatures of $\log T>$7.3. We further compare the DEM distributions sampled from the PS and nearby reference locations (Figure~\ref{fig:ps13-dem}g,h). The DEM profiles show that the PS samples have similar, cool EM components at \logt $<$6.7 to the references, but contain a much higher EM peak at \logt=6.9--7.3, providing further evidence that the PS feature in the flare contains mostly hot plasmas, and the cool DEM components are mainly due to the foreground and background emissions. Thus, the DEM results evidence a narrow temperature range of hot plasmas associated with the long sheet, i.e., \logt=6.9--7.3 for this case (PS\#1), which suggests a relatively isothermal feature.

\subsubsection{PS Length}

\begin{figure}[htbp]
	\centering
	\includegraphics[width=\textwidth]{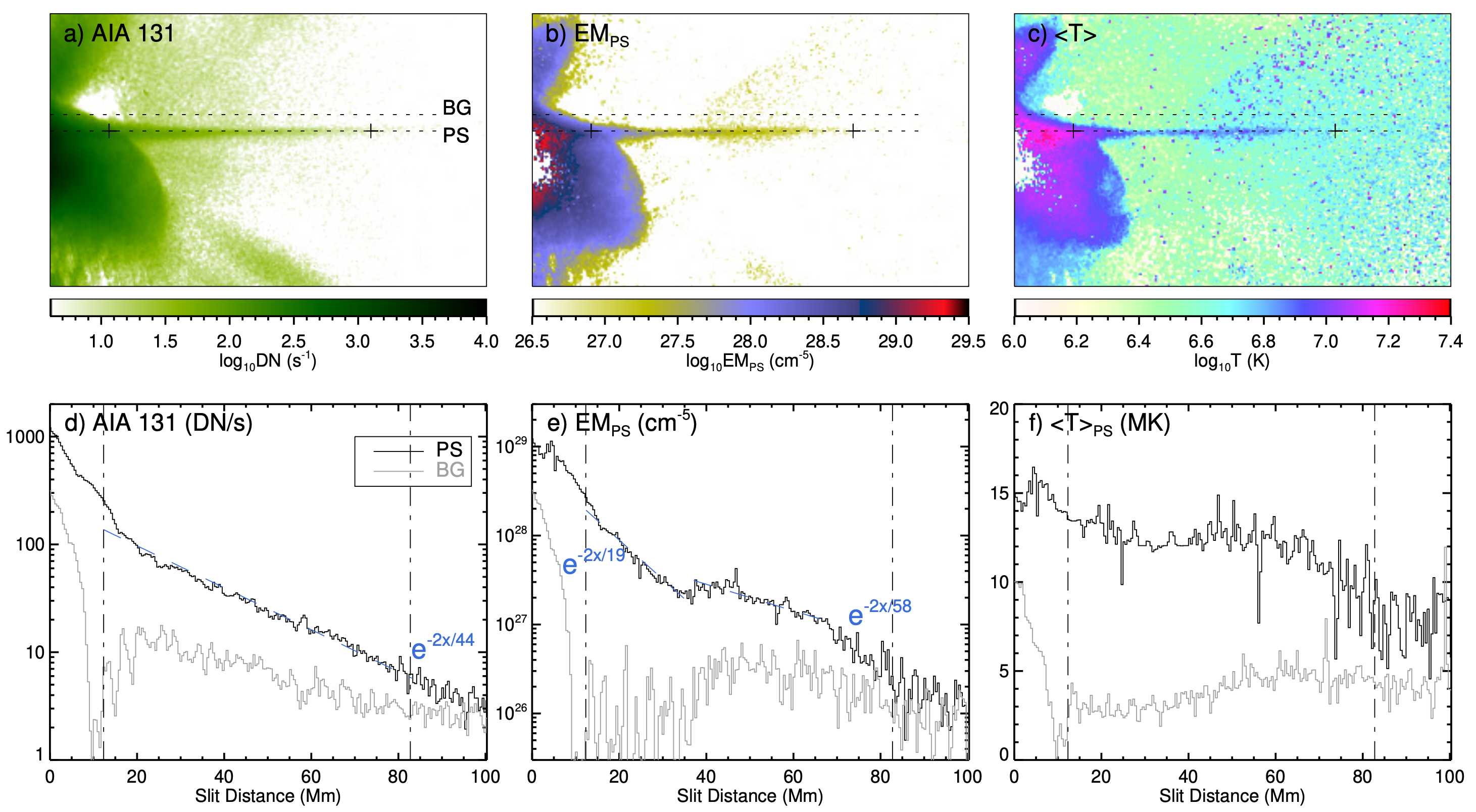}
	\caption{\small Distribution of AIA intensity, EM, and temperature along the PS structure during the 2013 May 13 flare. Top panels show maps of AIA 131 \AA, \emps, and the mean temperature \mt. The two horizontal dotted lines indicate the sample locations (averaged over 3 pixels) for the curves plotted in bottom panels, PS in black and BG (separated from PS by 6$\arcsec$) in gray, respectively. The blue dashed lines in panels (d,e) show fitted results to the profiles using exponential functions $\sim\mathrm{exp}(-2x/H)$, where $H$ gives the scale height of emissions. The temperature curves in panel (f) are from the mean temperatures weighted by only hot DEM component for the PS (\mtps; black) and by the whole temperature range for the BG (\mt, same as panel (c); gray), respectively. The two vertical dot-dashed lines in bottom panels mark the start and end of the observed PS (derived from Figure~\ref{fig:ps13-len-dem}), which are also indicated by two plus symbols in the top panels.}
	\label{fig:ps13-len-131}
\end{figure}

\begin{figure}[htbp]
	\centering
	\includegraphics[width=\textwidth]{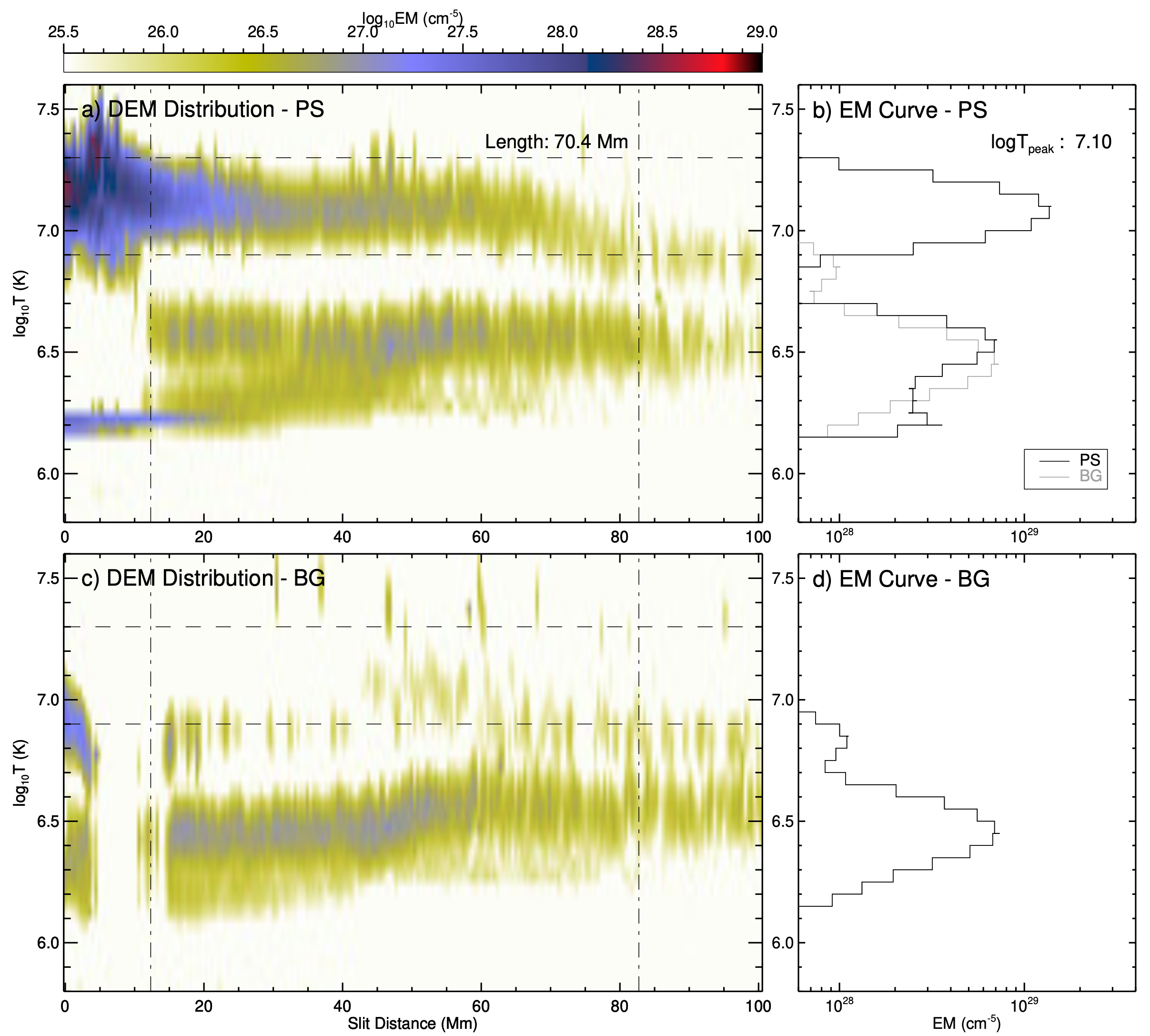}
	\caption{\small DEM distribution along the PS during the 2013 May 13 flare. Panels (a,c) show the DEM distributions from the PS and nearby background locations, indicated by the two parallel dotted lines in Figure~\ref{fig:ps13-len-131}(a--c), respectively. The two horizontal dashed lines mark the specific temperature range of PS, and two vertical dot-dashed lines mark the start and end of the hot PS plasmas, which gives the length of PS as labelled in panel (a). Panels (b,d) show the curves of total EMs summed along the PS's length, i.e., between the two vertical dot-dashed lines in panels (a,c), respectively. The peak temperature of the total EMs from PS is labelled in panel (b). The EM curve from BG in panel (d) is overplotted in panel (b) in gray for comparison.}
	\label{fig:ps13-len-dem}
\end{figure}

Based on the PS's temperatures determined from the DEM analysis, i.e., the narrow temperature range of the hot DEM component, we calculate the pure emission \emps and the mean temperature \mtps for the PS; for the whole region we calculate the conventional mean temperature \mt (Figure~\ref{fig:ps13-len-131}). The EM and mean temperature maps show a linear structure above the post-flare loops similar to the AIA 131~\AA\ observation. To investigate the PS properties at different heights, we plot the AIA intensity, \emps, and \mtps at a virtual slit along the PS structure; another parallel slit is placed nearby as a reference (the two dotted lines in Figure~\ref{fig:ps13-len-131}a--c). The AIA 131~\AA\ intensity is strong at the bottom of PS with emissions from overlapped flare loops, and then decreases smoothly with increasing heights, giving a scale height of $H$=44~Mm. The \emps profile decreases sharply at first, but reaches a plateau after $x\approx$35~Mm with the EM of $\sim 2\times10^{27}$~cm$^{-5}$, and finally drops to the background level of about $10^{26}$~cm$^{-5}$. We use two exponential functions to fit the two-episode EM variation separately, which exhibits a scale height of $\sim$20 and 60~Mm, respectively, suggestive of different density distributions at different heights. The temperature \mtps (the black curve in Figure~\ref{fig:ps13-len-131}f) is about 12--14~MK along the PS and is significantly higher than the conventional mean temperature \mt (about 6--8~MK) in Figure~\ref{fig:ps13-len-131}c, the latter of which underestimates the PS temperature by counting in significant cool foreground and background emissions along the line of sight \citep[see also][]{Gou2015}. For the reference BG, the missing emissions in AIA 131~\AA\ at around $x=$10~Mm, and the resultant zeros in EM and mean temperature (gray curves in Figure~\ref{fig:ps13-len-131}d--f; also Figure~\ref{fig:ps13-len-dem}c), are resulted from the removal of diffraction patterns by PSF deconvolving (Figure~\ref{fig:ps13-aia}).

To investigate the detailed temperature structure at different heights, we plot the DEM distribution over temperature as a function of distance along the PS (Figure~\ref{fig:ps13-len-dem}). Only the temperature range of \logt=5.8--7.6 is plotted in the figure as cool EMs from \logt$<$5.8 are much fewer (Figure~\ref{fig:ps13-dem}g,h). Comparing the DEM distributions from the PS and nearby BG, the cool EM components at \logt=6.1--6.7 are similar in magnitude, but the hot EM component only appears along the PS. Below the PS ($x\lesssim$12~Mm), the hot component contains EMs from a wider temperature range (Figure~\ref{fig:ps13-len-dem}a), which includes contributions from overlying flare loops and the cusp tip (Figure~\ref{fig:ps13-dem}c,f). 

The most interesting feature is that the temperature almost keeps the same along the PS, i.e., EMs are all from the narrow temperature range of \logt=6.9--7.3 (between the two horizontal dashed lines in Figure~\ref{fig:ps13-len-dem}a), although at $x\gtrsim$70~Mm the temperature drops a little where the AIA intensity is low at the upper tip. According to the well-defined temperature range of hot plasmas, we obtain the length of PS, which is about 70~Mm as indicated by two vertical lines in Figure~\ref{fig:ps13-len-dem}a. The measured length from the DEM distribution agrees with the variations of AIA intensity and \emps along the PS, which are distinctly higher than the BG level (Figure~\ref{fig:ps13-len-131}d,e). Considering that the upper part could be too faint to observe, the real length of PS will be longer. By summing up the total EMs along the whole length of the PS structure (within the two vertical lines in Figure~\ref{fig:ps13-len-dem}a), we obtain a peak temperature of \logt=7.1 (about 12.6~MK; Figure~\ref{fig:ps13-len-dem}b) for the PS, in agreement with \mtps in Figure~\ref{fig:ps13-len-131}f. The \emps decreases from about 3$\times10^{28}$ to 2$\times10^{26}~\mathrm{cm}^{-5}$ along the PS, which corresponds to a plasma density of about 1.6$\times10^{9}$--1.3$\times10^{8}~\mathrm{cm}^{-3}$, assuming the line-of-sight depth of $\sim$120~Mm based on stereoscopic observations from two satellites \citep[see, e.g., Fig.3 in][]{Gou2019}. 

\subsubsection{PS Thickness}

\begin{figure}[htbp]
	\centering
	\includegraphics[width=\textwidth]{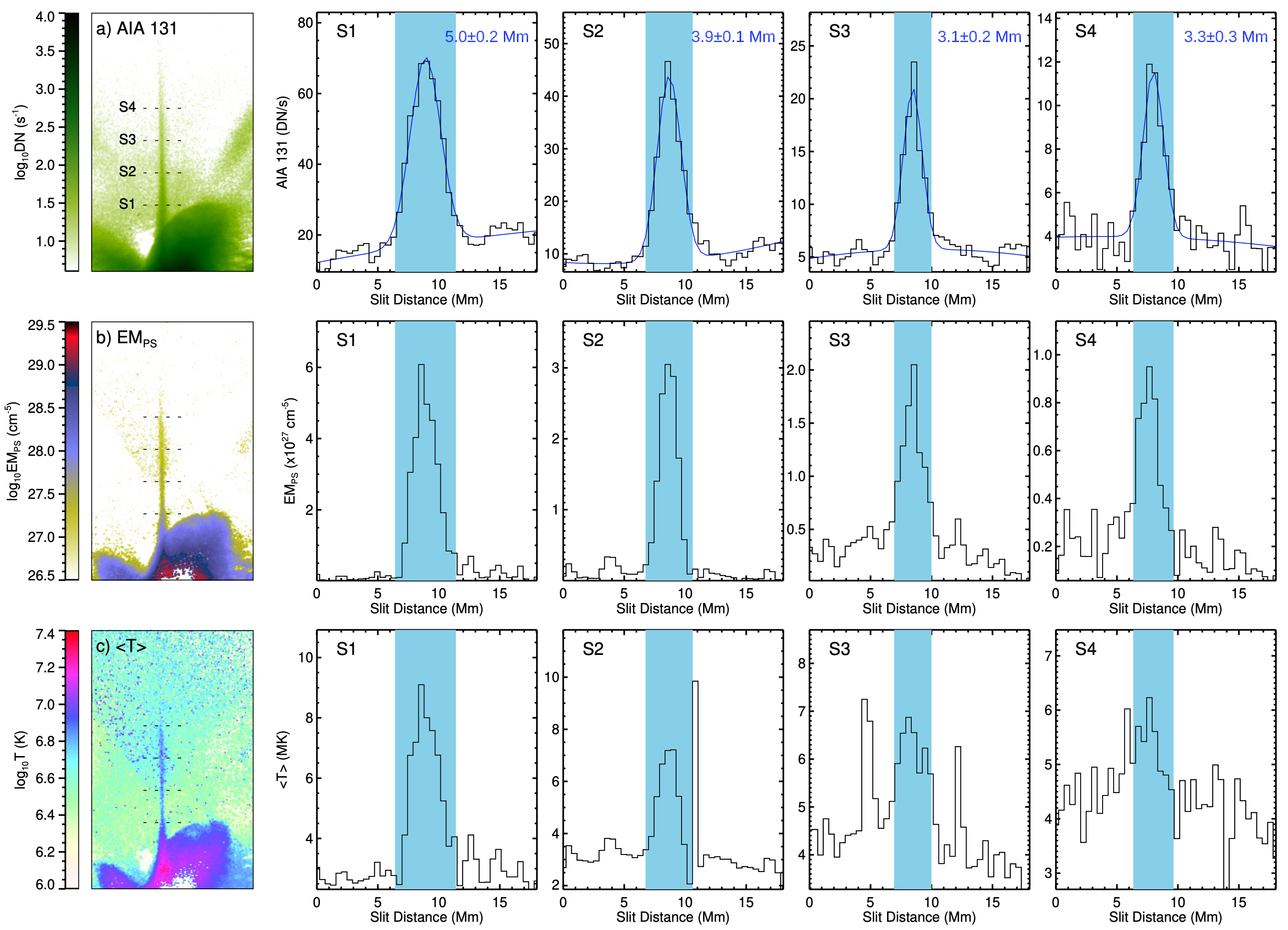}
	\caption{\small Distributions of the AIA intensity, EM, and temperature across the PS structure. Panels on the left show maps of AIA 131 \AA, \emps, and the mean temperature \mt. The four dotted lines indicate sample locations (S1--S4; averaged over 3 pixels) to generate the curves plotted in the right four panels, respectively, which are parallel placed and separated from each other by 20$\arcsec$. The blue curves in top panels are the fitting results of the AIA 131~\AA\ intensities, and the light blue shades mark 2$\sigma$ widths of the Gaussian fittings. The width values and uncertainties of fits are labelled in blue at the top-right corner in each panel. The widths of Gaussian fittings to the AIA intensities are also overplotted on the \emps and temperature profiles in the middle and bottom panels for comparison.}
	\label{fig:ps13-thi-131}
\end{figure}

\begin{figure}[htbp]
	\centering
	\includegraphics[width=\textwidth]{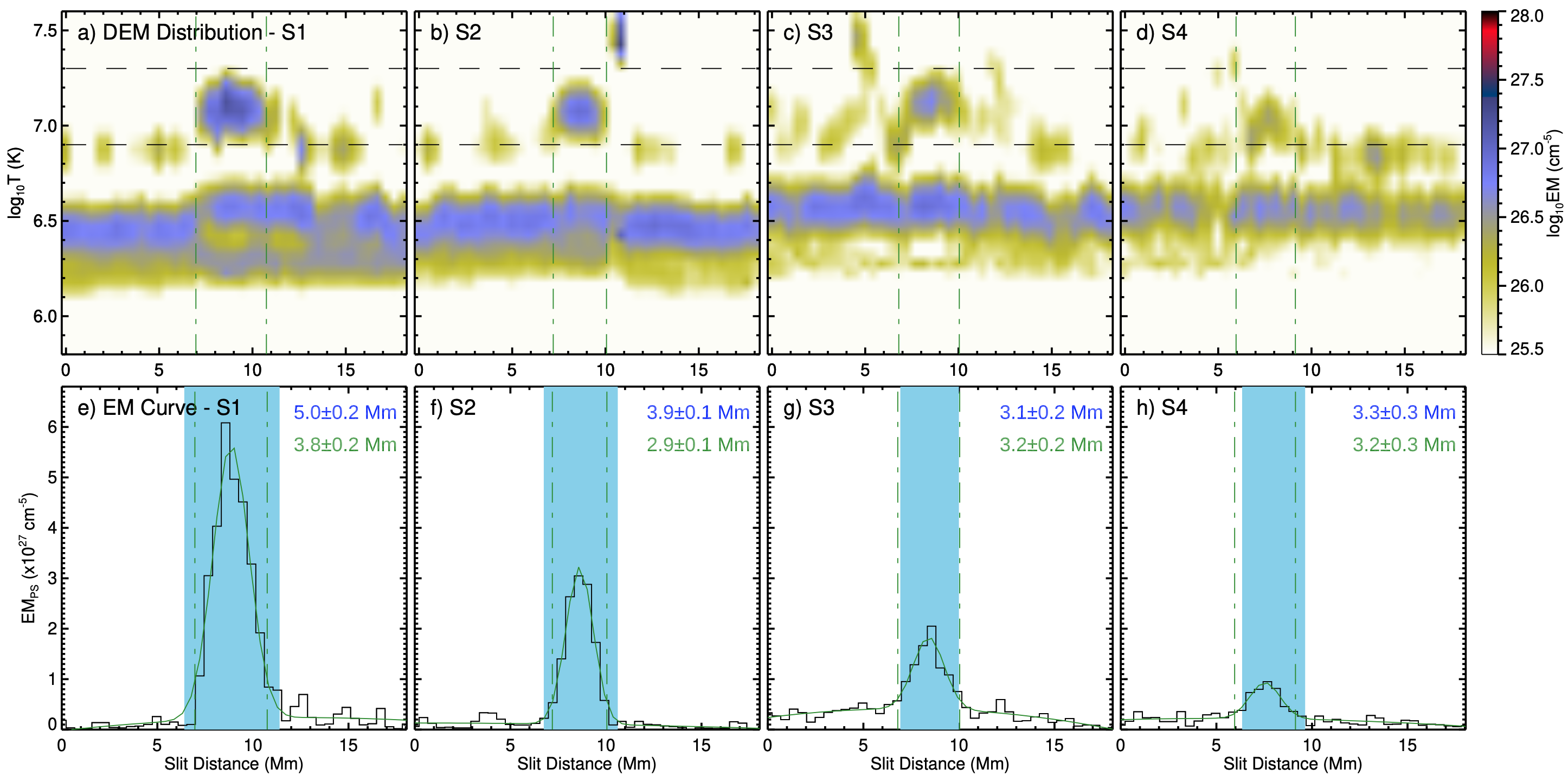}
	\caption{\small DEM distribution across the PS. Top panels show the DEM distributions from the sample locations S1--S4 in Figure~\ref{fig:ps13-thi-131}. The two horizontal dashed lines mark the temperature range of PS (the same as those in Figure~\ref{fig:ps13-len-dem}a,c). Bottom panels show distributions of \emps along S1--S4, i.e., the EM profiles summed over DEMs between the two horizontal dashed lines in top panels. The solid green curves in bottom panels are the fitting results of the \emps profiles, and the two vertical dot-dashed lines in green mark 2$\sigma$ widths of the Gaussian fittings, with values and errors labelled in green. The two vertical green lines are also plotted on the DEM distributions in top panels, which are consistent with the location of hot plasmas from PS. The light blue shades and widths in blue in bottom panels are taken from those in Figure~\ref{fig:ps13-thi-131} for comparison.}
	\label{fig:ps13-thi-dem}
\end{figure}

We also investigate the apparent thickness of the PS structure. We place four parallel slits at different heights across the PS and plot the profiles of AIA 131~\AA\ intensity, \emps, and mean temperature \mt (Figure~\ref{fig:ps13-thi-131}). The PS is clearly associated with enhanced emissions and temperatures similar to a Gaussian shape. To measure the PS thickness observed in AIA 131~\AA, we use a Gaussian function plus a second-order polynomial to fit the intensity profile, the latter of which indicates background emissions off the PS \citep[see also][]{Warren2018}. We use 2$\sigma$ width of the Gaussian fit as the thickness of PS (blue shaded regions in Figure~\ref{fig:ps13-thi-131}), where the intensity is significantly higher than the background. The PS thickness observed in AIA 131~\AA\ is about 3--5 Mm in this case, and the value is generally smaller at a larger height.

We plot the DEM distribution over temperature as a function of distance across the PS (Figure~\ref{fig:ps13-thi-dem}). We can always see the background EM component at temperatures of \logt =6.1--6.7, which changes little along the slits. The most prominent feature is the narrow EM component in the temperature range of PS, \logt =6.9--7.3, whose two edges give a good estimation of the PS thickness by containing only hot plasmas. We similarly use a Gaussian function plus a second-order polynomial to fit the \emps curve and obtain the 2$\sigma$ width of Gaussian fitting (in green in Figure~\ref{fig:ps13-thi-dem}), where the fitted width is consistent with the location of the narrow, hot EM component from the PS (Figure~\ref{fig:ps13-thi-dem}a--d). The thickness measured from DEM results is about 3--4~Mm, slightly smaller than those from AIA 131~\AA\ intensity, and shows small differences for different heights.

\subsection{2017 September 10 X8.2 flare (PS\#2)}

 \begin{figure}[htbp]
	\centering
	\includegraphics[width=\textwidth]{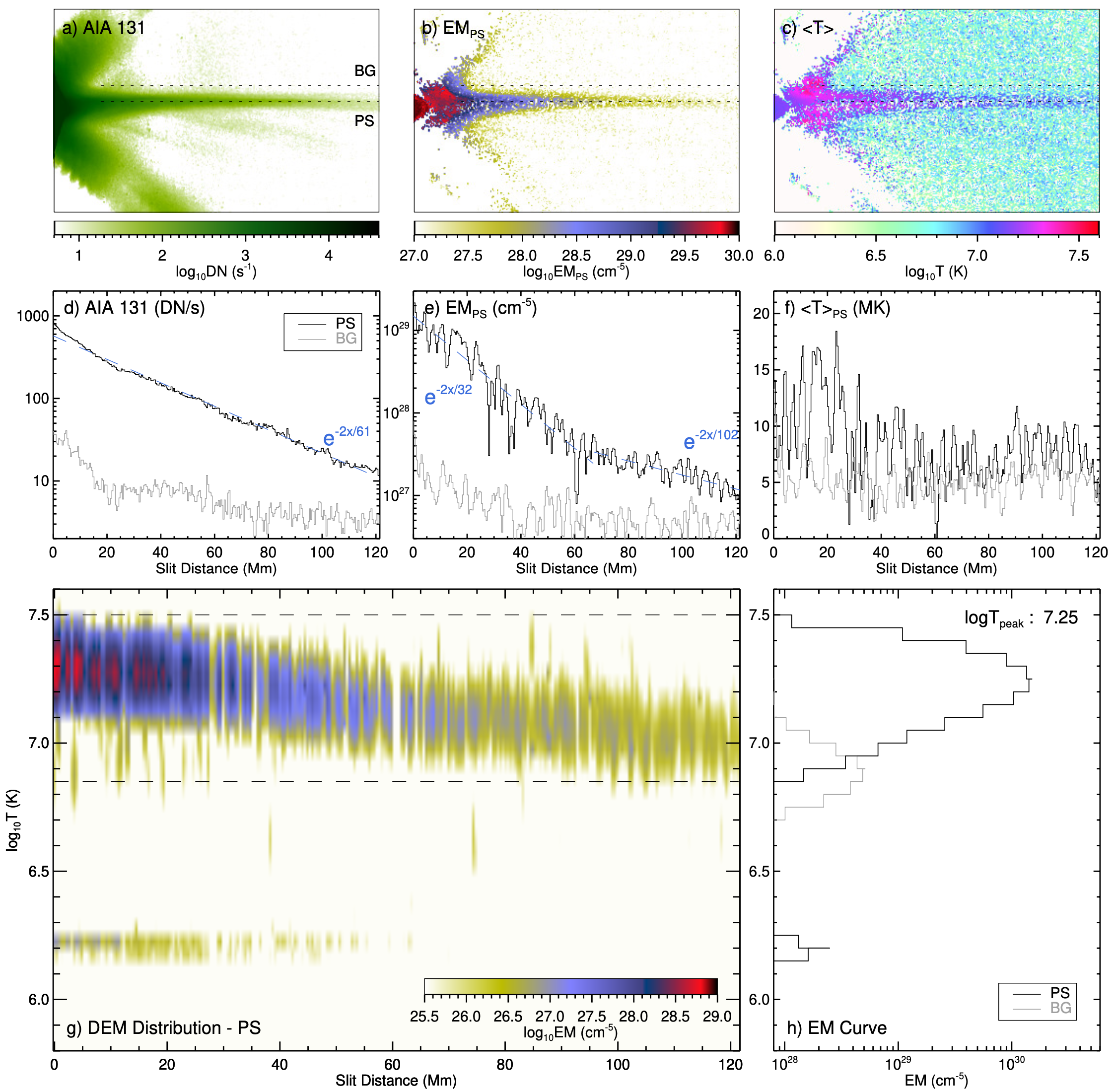}
	\caption{\small Distributions of AIA intensity, \emps, mean temperature, and the DEM along the PS during the 2017 September 10 flare. Panels (a--f) are the same as Figure~\ref{fig:ps13-len-131} but for PS\#2, and panels (g,h) are the same as Figure~\ref{fig:ps13-len-dem}a,b. The location of slit BG is separated from PS by 10$\arcsec$ in this case.}
	\label{fig:ps17-len}
\end{figure}

\begin{figure}[htbp]
	\centering
	\includegraphics[width=\textwidth]{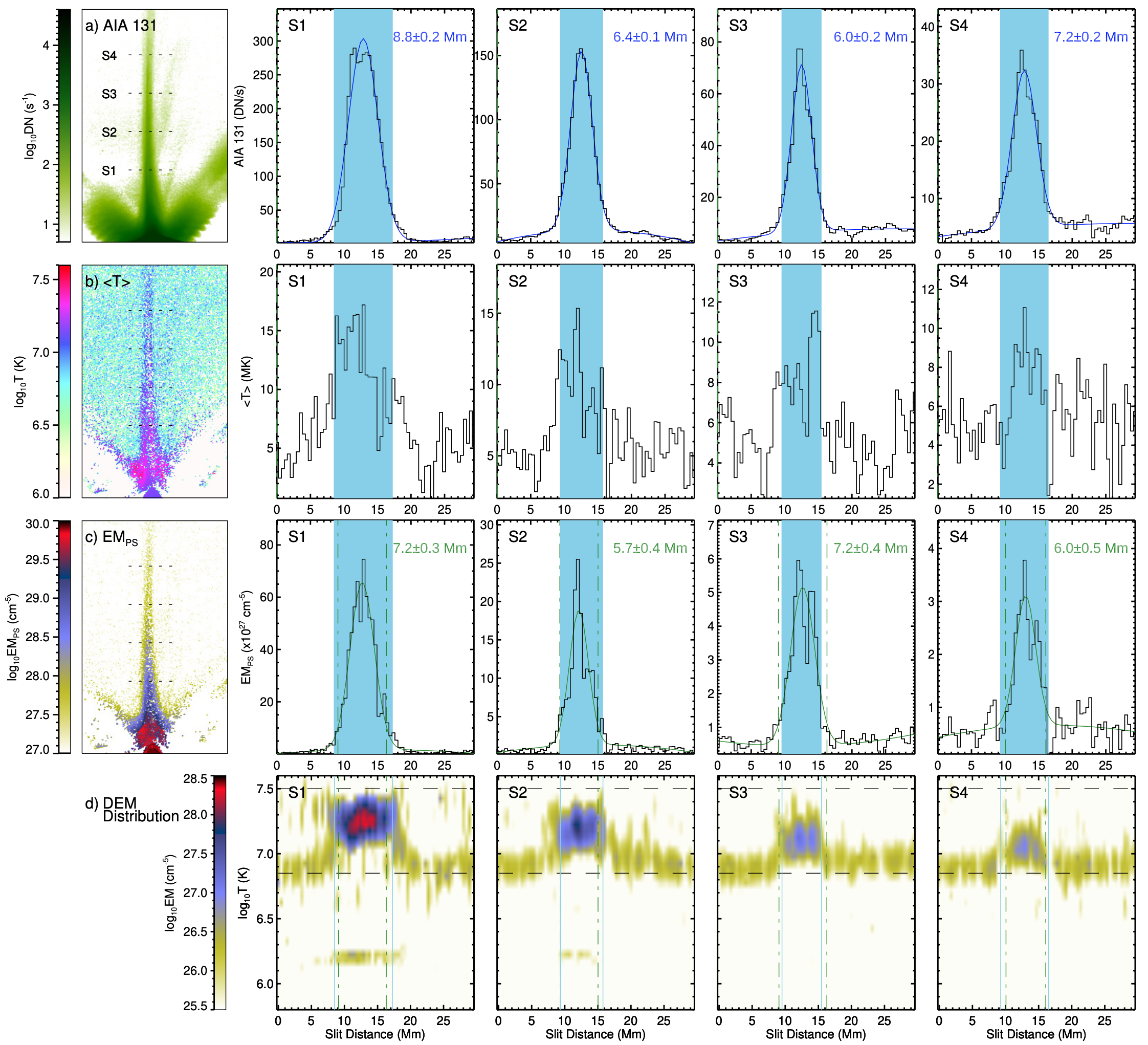}
	\caption{\small Distributions of the AIA intensity, mean temperature, \emps, and the DEM across the PS. Same as Figures~\ref{fig:ps13-thi-131}\&\ref{fig:ps13-thi-dem} but for PS\#2. The four parallel slits S1--S4 are separated by 33$\arcsec$ in this case. The fitted results to the AIA intensities and \emps curves are plotted in blue and green, respectively. The vertical light blue lines in bottom panels mark edges of the light blue shades from fittings to the AIA 131~\AA\ intensities. The vertical dot-dashed green lines are from fittings to the \emps profiles.}
	\label{fig:ps17-thi}
\end{figure}

The X8.2 flare on 2017 September 10 occurs in AR 12673 above the solar west limb, and exhibits a `textbook' eruption bearing a striking resemblance to the standard picture. A long sheet structure forms in the wake of an erupting flux rope, and the formation and dynamics of the PS feature have been studied in a number of papers using various observations from different instruments, including multi-wavelength imaging, and EUV and radio spectroscopies \citep[e.g.,][]{Seaton2018,LiY2018,Warren2018,Cheng2018,Longcope2018,French2019,Chen2020}. The PS structure during the flare gradual phase is observed to extend beyond the AIA's FOV \citep{Seaton2018}, and it is visible in all six EUV channels of AIA due to continuum emissions \citep[see, e.g., Figure 7 in][]{Warren2018}. We use DEM analysis to study the PS structure observed at 16:41~UT and show the results in Figures~\ref{fig:ps17-len}\&\ref{fig:ps17-thi}.

We similarly place two parallel virtual slits on the PS structure and its nearby location, respectively, to examine the temperature structure and its distribution with height (Figures~\ref{fig:ps17-len}). We skip the saturated flare loops underneath and the slit is completely on the PS structure itself, which has a length of $>$120~Mm in AIA's FOV. The DEM distribution of PS shows a single hot component from temperatures of \logt = 6.85--7.5, which we use as the temperature range of PS\#2, and contains few emissions from lower temperatures (Figure~\ref{fig:ps17-len}g,h); while the nearby BG shows a cooler EM component from \logt= 6.7--7.1 (the gray curve in Figure~\ref{fig:ps17-len}h). The DEM distribution is consistent with the results in \citet{Warren2018} (e.g., Figure~8) using a different DEM inversion method. The narrow temperature range of EM suggests an isothermal feature of the plasmas surrounding the flare current sheet. 

By examining the DEM distribution at different heights along the PS, one can see that the plasma temperatures decrease slightly from \logt = 7.1--7.5 to 6.85--7.2 with increasing heights above the flare arcade (Figure~\ref{fig:ps17-len}g). This result is different from the 2013 May 13 flare that show constant temperatures at different heights (PS\#1 in Figure~\ref{fig:ps13-len-dem}). The integration of EMs along the CS gives a peak temperature of \logt=7.25 for the plasmas, which is generally hotter than in the previous case (PS\#1). The \emps changes between 2$\times10^{29}$ and 10$^{27}~\mathrm{cm}^{-5}$ (Figure~\ref{fig:ps17-len}e), which is one magnitude denser than the PS\#1. The \emps distribution at different heights first decreases sharply following an exponential scale height of $H$=32~Mm, while the second part (e.g., $x>70$~Mm) shows a much slower descending with increasing heights ($H$=102~Mm). The fast-to-slow descending trend in EM agrees with the results in \citet{Longcope2018} of the same event, but different from that in AIA 131~\AA\ intensity, which decreases smoothly all the way along the PS (Figure~\ref{fig:ps17-len}d). 

We further measure the PS's thickness by placing four parallel slits separating in heights across the PS and fitting the AIA and \emps profiles (Figure~\ref{fig:ps17-thi}). The DEM distribution shows constant background emissions at \logt=6.75--7.0, and hotter emissions only from the PS feature; the latter shows decreasing temperatures with increasing heights. The thickness of PS is measured as about 6--9~Mm in AIA 131~\AA\ intensity, and about 6--7~Mm in DEM, which is twice of the value in the previous case. For comparison, \citet{Cheng2018} measured the PS's thickness as $\sim$10~Mm based on the total EMs and \citet{LiY2018} measured as 7--11~Mm from the spectral line intensity and non-thermal broadening for the same flare at different times.

\subsection{2023 February 17 X2.3 Flare (PS\#3)}

\begin{figure}[htbp]
	\centering
	\includegraphics[width=\textwidth]{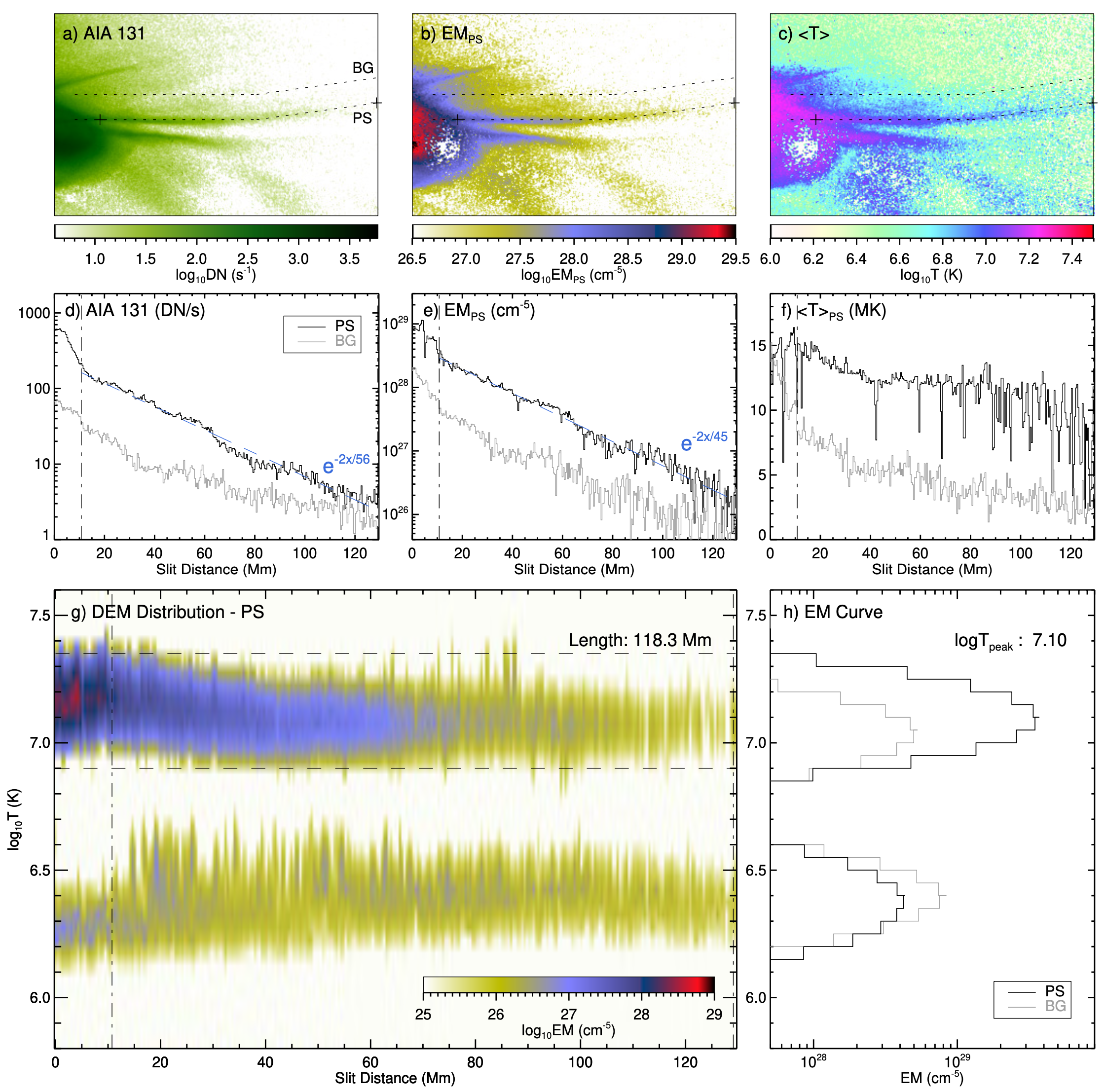}
	\caption{\small Properties along the PS structure during the 2023 February 17 flare. Same as Figure~\ref{fig:ps17-len} but for the PS\#3. Two parallel curved slits are used to sample PS and BG in this case, which are separated by 15$\arcsec$.}
	\label{fig:ps23-len}
\end{figure}

\begin{figure}[htbp]
	\centering
	\includegraphics[width=\textwidth]{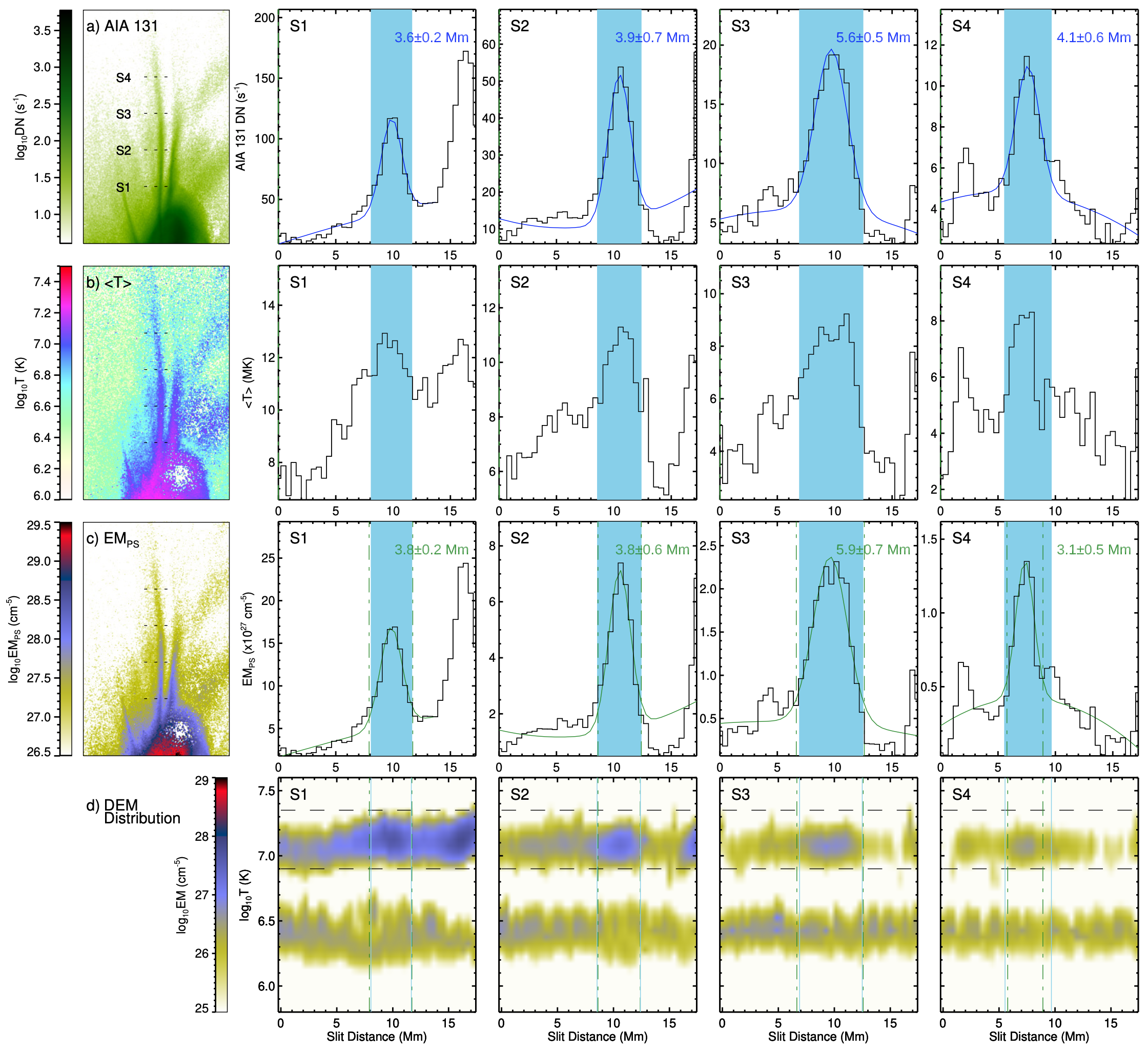}
	\caption{\small Properties across the PS structure. Same as Figure~\ref{fig:ps17-thi} but for PS\#3. S1--S4 are separated by 30$\arcsec$ in this case.}
	\label{fig:ps23-thi}
\end{figure}

The X2.3 flare on 2023 February 17 occurs in AR 13229 near the solar northeastern limb. The observational perspective of AIA is not perfectly along the axis of the flare arcade but has a slight tilt, thus gives a different morphology of PS from an ideal linear feature. During the flare gradual phase, the emissions above the flare arcade in AIA 131~\AA\ exhibit several bright spikes surrounded by less bright plasmas (PS\#3 in Figure~\ref{fig:flares}), forming a supra-arcade fan similar to those viewed face-on. Therefore, this event provides a complement for other edge-on cases. We focus on one of the longest spikes observed at 20:45~UT, which serves as a good representative for the PS feature above flare loops.

We perform DEM analysis for this PS case and plot the results in Figures~\ref{fig:ps23-len} \&\ \ref{fig:ps23-thi}. Since the structure is curved a little as it extends upward, we use two parallel curved slits to sample the PS and its nearby reference location. The DEM distribution from the PS shows two EM components from temperatures of \logt = 6.1--6.6 and 6.9--7.4, respectively, which generally agrees with that from the BG but shows a much greater hot EM component than BG (Figure~\ref{fig:ps23-len}g,h). This high-temperature enhancement is also evidenced in the DEM distributions from the sample locations across the PS (Figure~\ref{fig:ps23-thi}d), which show two EM components with similar temperatures all across the PS but having more hot emissions in the middle. The DEM distributions suggest that the plasma structure above the flare arcade holds similar temperatures, and the bright spike contains dense hot plasmas from temperatures of \logt=6.9--7.4. 

Another notable feature is that the temperatures are similar at different heights along the PS (Figure~\ref{fig:ps23-len}g), which is consistent with the 2013 flare case (PS\#1; Figure~\ref{fig:ps13-len-dem}) but not the 2017 one (PS\#2; Figure~\ref{fig:ps17-len}). For the PS's length in this case, the lower point of the structure is difficult to determine as it overlaps on the flare loop-top with an oblique viewing angle and the temperatures are also similar. We generally select a sudden drop in the \emps curve as the start of PS and measure the observed length in AIA as about 118~Mm (Figure~\ref{fig:ps23-len}). The total EM along the PS exhibits a peak temperature of \logt=7.1, in agreement with \mtps (Figure~\ref{fig:ps23-len}f,h). Unlike the temperature, the AIA 131~\AA\ intensity and \emps along the PS descend intensively. For this case, the \emps falls over height with a smooth trend similar to the AIA 131~\AA\ intensity, but different from the fast-to-slow descending in PSs~\#1\&2. The hot EMs from PS are in the same magnitude with PS\#1. The PS's thickness, measured by placing four parallel slits across the structure, is about 3--6~Mm (Figure~\ref{fig:ps23-thi}).

\subsection{2014 February 25 and 2013 May 14 Flares}

\begin{figure}[htbp]
	\centering
	\includegraphics[width=\textwidth]{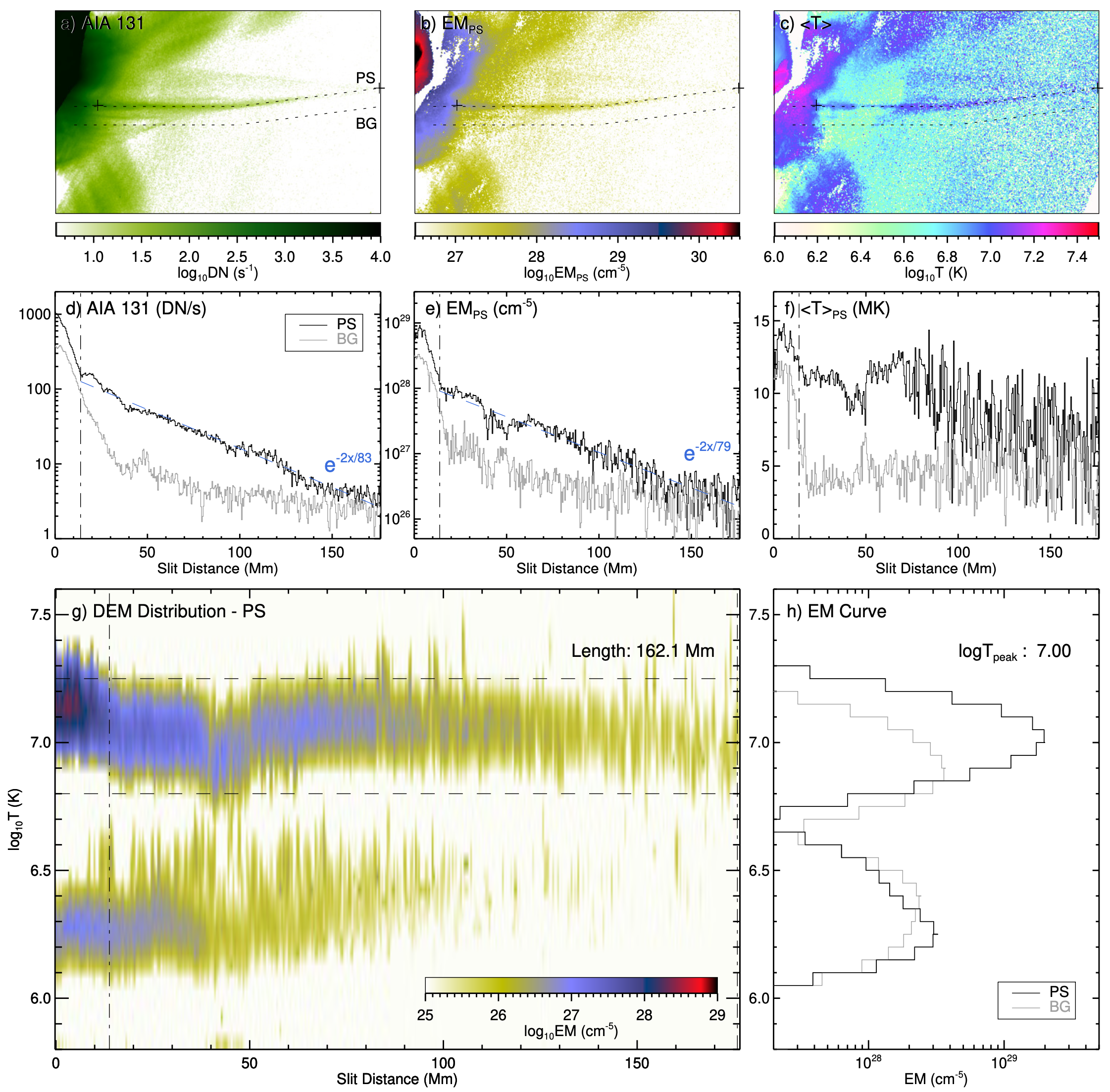}
	\caption{\small Properties along the PS structure during the 2014 February 25 flare. Same as previous figures but for the PS\#4. The two parallel curved slits are separated by 15$\arcsec$.}
	\label{fig:ps14-len}
\end{figure}

\begin{figure}[htbp]
	\centering
	\includegraphics[width=\textwidth]{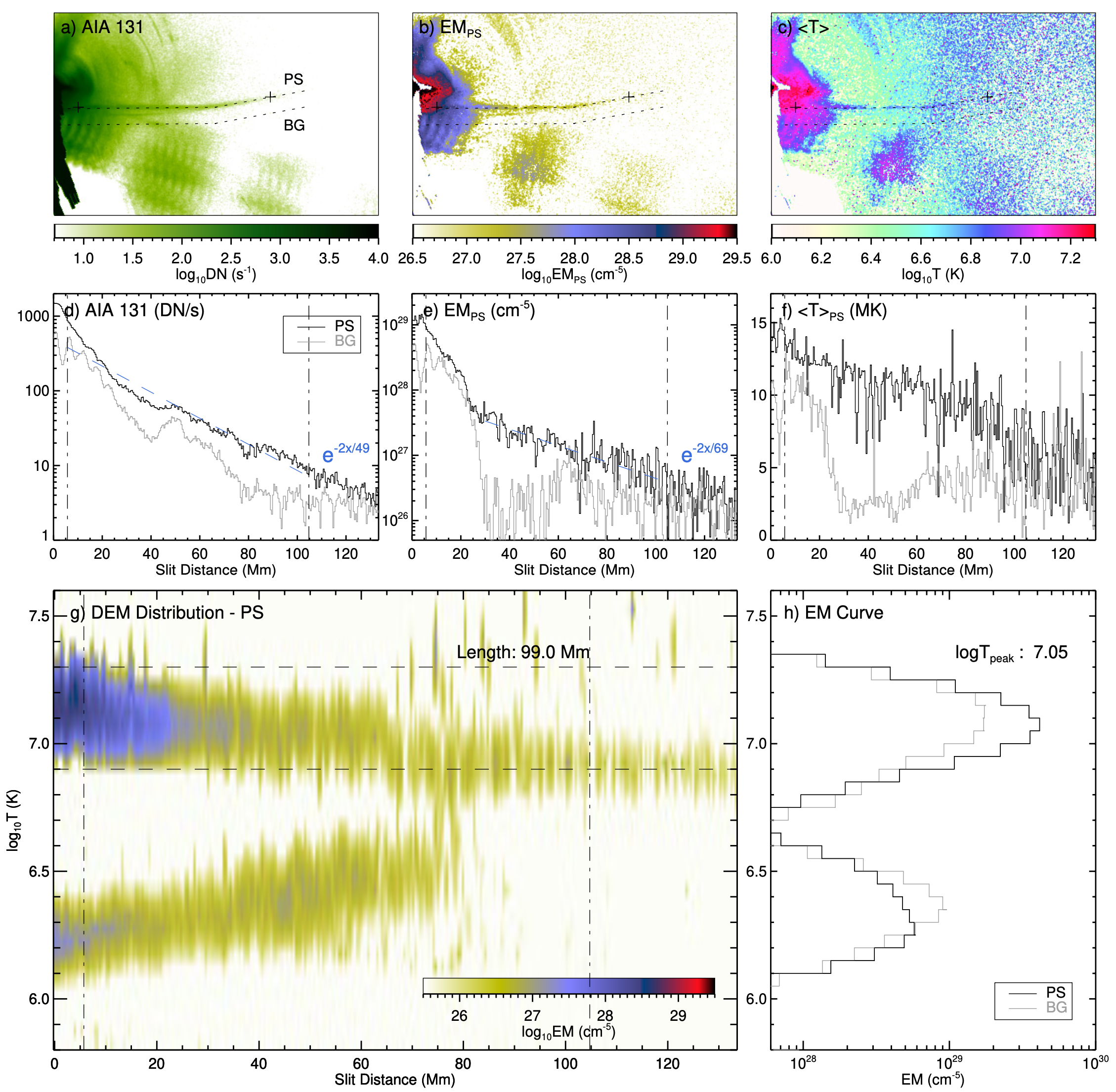}
	\caption{\small Properties along the PS structure during the 2013 May 14 flare. Same as previous figures but for the PS\#5. The two parallel curved slits are separated by 12$\arcsec$.}
	\label{fig:ps13b-len}
\end{figure}

We study two more flares which show a long PS feature during the gradual phase. The X4.9 flare on 2014 February 25 occurs near the solar southeast limb (PS\#4; Figure~\ref{fig:flares}). Due to a slightly oblique view angle, the sheet structure above the flare arcade exhibits as a three-dimensional (3D) feature with some faint emissions on the either side of a linear structure. The formation and development of PS in the flare was studied in detail in \citet{Seaton2017}. We investigate the long PS feature observed at 01:08~UT shortly after the flare peak. The DEM analysis shows that the emissions of PS are from temperatures at \logt=6.8--7.25 with a peak at \logt=7.0, and the temperatures generally remain constant at different heights along the PS (Figure~\ref{fig:ps14-len}). The \emps and AIA intensity fall over height following similar scales. The measured length is about 162~Mm (Figure~\ref{fig:ps14-len}) and the thickness is about 2--6~Mm (We didn't show the figures of thickness measurements for this case and following ones, but all measurements are presented in Section~\ref{sec:discussion}).

The X3.2 flare on 2023 May 14 occurs near the solar northeast limb in AR 11748 (PS\#5; Figure~\ref{fig:flares}), in the same AR with the 2013 May 13 flare (PS\#1) but several hours later. The flare exhibits a linear PS feature above post-flare arcade during the gradual phase as well, although the view angle is a little more oblique than PS\#1. For this case, the diffraction pattern and overlapping loops have an impact on the PS feature, but its properties are still accessible. The DEM analysis shows that the temperature of PS is in the narrow range of \logt=6.9--7.3 and peaks at \logt=7.05 (Figure~\ref{fig:ps13b-len}). The temperature is also similar at different heights along the PS, while the upper part seems too faint to be resolved. Since the bottom of PS overlaps largely on flare loops, we only fit the upper portion of the \emps curve, which holds a scale height of $\sim$70~Mm. We obtain the PS's length of about 99~Mm and the thickness of about 2--5~Mm.

\subsection{PSs During Flare Impulsive Phase}

\begin{figure}[htbp]
	\centering
	\includegraphics[width=\textwidth]{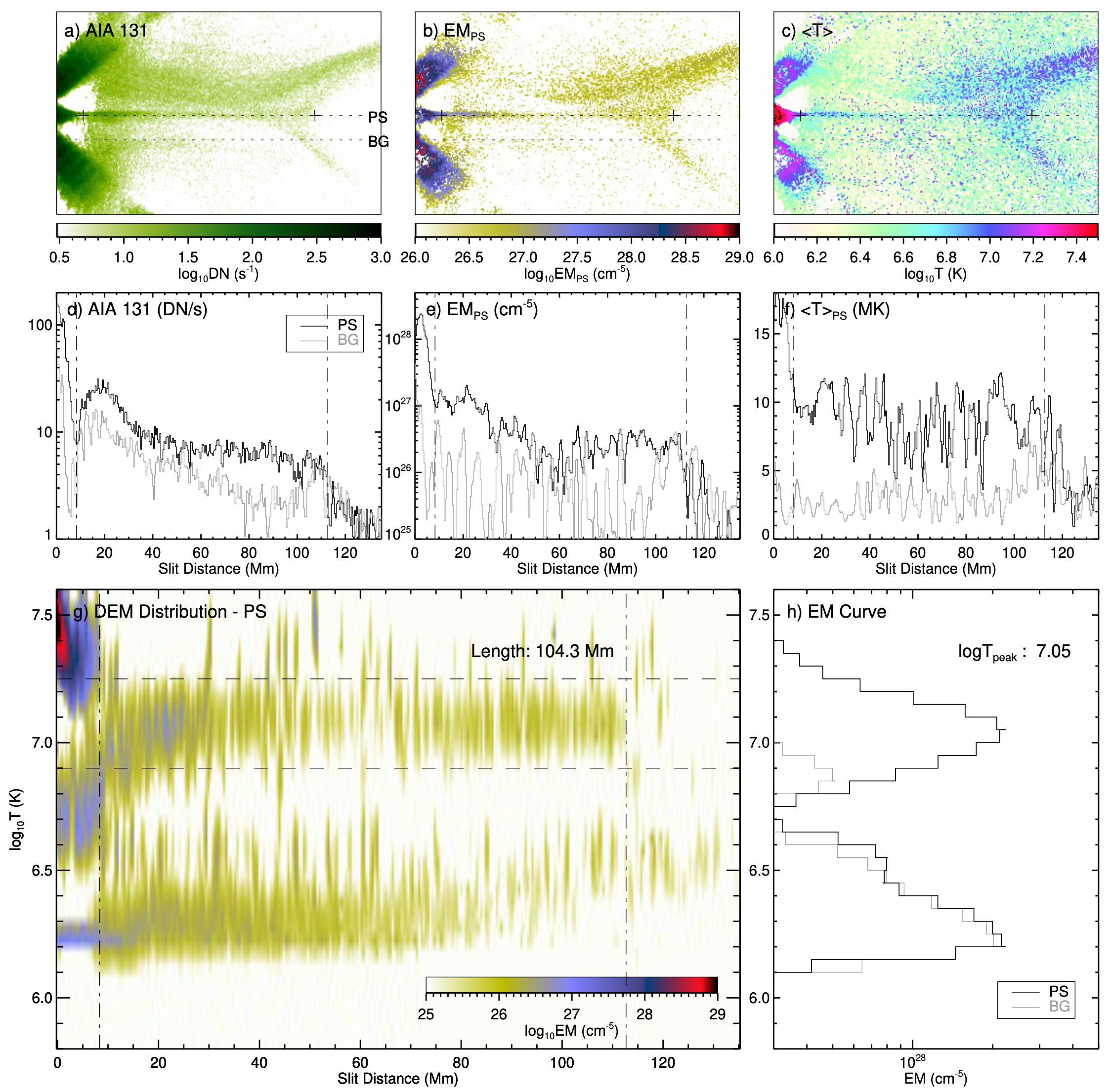}
	\caption{\small Properties along the PS structure during the impulsive phase of 2017 September 10 flare (PS\#6). The two parallel slits are separated by 10$\arcsec$.}
	\label{fig:ps17i-len}
\end{figure}

\begin{figure}[htbp]
	\centering
	\includegraphics[width=\textwidth]{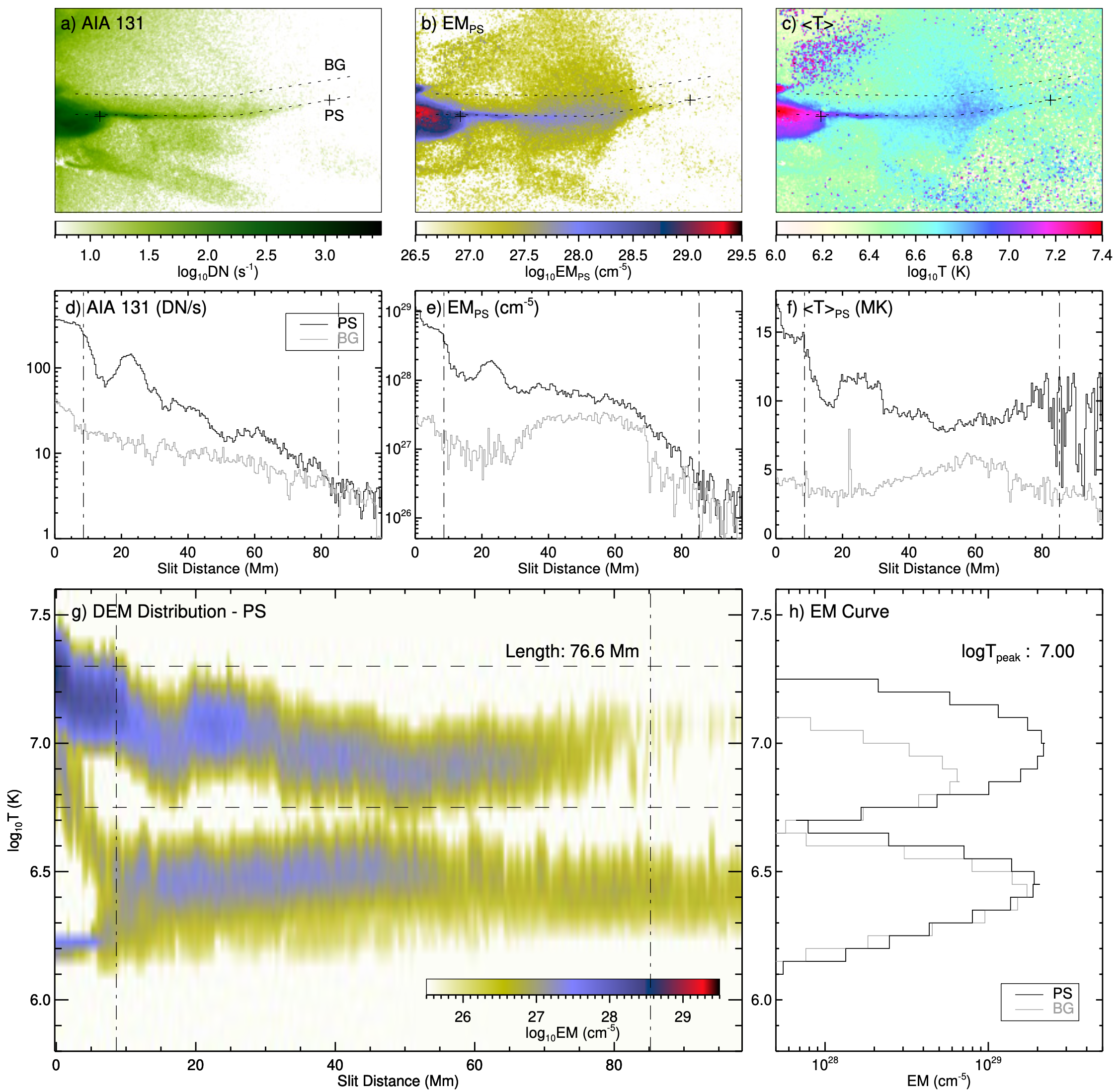}
	\caption{\small Properties along the PS structure during the impulsive phase of 2023 February 17 flare (PS\#7). The two parallel curved slits are separated by 10$\arcsec$.}
	\label{fig:ps23i-len}
\end{figure}

We study the PS feature observed during the impulsive phase of flares and compare with that during the gradual phase. The 2017 September 10 flare during the impulsive eruption exhibits a thin, linear PS feature in AIA 131~\AA, which connects the bottom of an erupting flux rope and the tip of cusp-shaped flare loops underneath, resembling the standard flare model (PS\#6; Figure~\ref{fig:flares}). We analyze the PS structure observed at 15:55~UT when the trailing part of the erupting flux rope, an inverted-Y shape, is still visible in the AIA's FOV. We obtain a narrow temperature range of \logt=6.9--7.25 and a peak temperature of \logt=7.05 for the PS (Figure~\ref{fig:ps17i-len}), which is generally cooler than that during the gradual phase of the same flare (PS\#2; Figure~\ref{fig:ps17-len}). The temperatures for this case remain almost constant at different heights along the PS, which is also different from PS\#2 during the gradual phase showing decreasing temperatures with increasing heights. The PS's length measured between the cusp tip of flare loops and the bottom of inverted-Y shape is about 104~Mm, which gives a complete length of the current sheet region. The PS's thickness is about 1--4~Mm, significantly thinner than during the gradual phase. The EMs for the PS are between $\sim4\times10^{25}$ and 2$\times10^{27}$~cm$^{-5}$, which is about two orders smaller in magnitude than PS\#2. Unlike the gradual phase cases, the EM and AIA 131~\AA\ intensity during impulsive phase do not show an exponential drop-off over height, but exhibit fluctuations while descending slowly, suggestive of nonuniform plasma densities along the sheet (Figure~\ref{fig:ps17i-len}d,e).

During the impulsive phase of the 2023 February 17 flare, a thin PS feature is observed in AIA 131~\AA\ in the wake of the eruption of a plasma cloud and connects to the cusp-shaped flare loop system underneath (PS\#7; Figure~\ref{fig:flares}). We study the PS structure observed at 20:06~UT before the peak in SXR flux of the flare. The DEM distribution shows PS emissions from temperatures at \logt=6.75--7.3 with a peak of at \logt=7.0 (Figure~\ref{fig:ps23i-len}), which is slightly cooler than that during the gradual phase of the same flare (PS\#3; Figure~\ref{fig:ps23-len}). The surrounding locations also contain a hot EM component from temperatures of \logt=6.7--7.1 (the gray curve in Figure~\ref{fig:ps23i-len}h). The temperatures are generally similar at different heights along the PS, with a slight enhancement near the bottom. We measure the length as about 77~Mm and a thickness of about 1--3~Mm, showing that the plasma sheet is thinner than during the gradual phase. The EMs are about $10^{26}$--$10^{28}$~cm$^{-5}$, which is similar to the gradual phase case but could also include the contribution from a cloud of hot plasmas surrounding the PS feature (see e.g., Figure~\ref{fig:ps23i-len}b). Fluctuations are clearly seen in \emps and AIA intensity profiles. For example, a notable peak occurs at the slit position of $x$=15-30~Mm in both AIA and \emps (Figure~\ref{fig:ps23i-len}d,e), which is also associated with an enhancement in temperature (Figure~\ref{fig:ps23i-len}f,g).

\section{Discussion} \label{sec:discussion}

\begin{table}[htb]
	\caption{Summary of measurements.} \label{tab:summary}  
    \centerline{
    \resizebox{1.1\textwidth}{!}{
	\begin{tabular}{cccccccccccc} 
		\hline\hline
		Flare & Phase & Case & \logt(K) & $\log T_{\rm peak}$(K) & \emps (cm$^{-5}$) &\multicolumn{2}{c}{Scale Height (Mm)} & Length (Mm) & \multicolumn{2}{c}{Thickness (Mm)} & Length-to-Thickness \\
        &&&&&&in AIA&in DEM&&in AIA\footnote{The PS thickness measured in AIA 131~\AA\ intensities. This column shows the mean value of thickness measured at four different heights and their standard deviation.} &in DEM\footnote{Same as the previous column but measured in DEM results, i.e., \emps profiles.}&Ratio\footnote{Minimum of the length-to-thickness ratio of PS, divided by the larger thickness value shown in the left two columns.} \\
		\hline\hline
        2013-05-13 & Gradual & PS\#1 & 6.90 -- 7.30 & 7.10 & 
        2.7$\times10^{28}$ -- 1.8$\times10^{26}$ &44&19, 58& $>$70.4 & 
        3.8$\pm$0.9 & 3.3$\pm$0.4 & $>$18.5 \\[1ex]
		\hline
        2013-05-14 & Gradual & PS\#5 & 6.90 -- 7.30 & 7.05 & 
        9.4$\times10^{28}$ -- 2.1$\times10^{26}$ &49&69& $>$99.0 &
        4.2$\pm$0.5 & 3.1$\pm$0.8 & $>$23.6 \\[1ex]
		\hline
		2014-02-25 & Gradual & PS\#4 & 6.80 -- 7.25 & 7.00 & 
        1.2$\times10^{28}$ -- 2.8$\times10^{26}$ &83&79& $>$162.1 &
        4.9$\pm$0.9 & 4.3$\pm$1.5  & $>$33.1 \\[1ex]
		\hline
		\multirow{2}{*}{2017-09-10} & Impulsive & PS\#6 & 6.90 -- 7.25 & 7.05 &
        2.0$\times10^{27}$ -- 4.4$\times10^{25}$ &N/A&N/A& $\approx$104.3 &
        2.6$\pm$0.8 & 1.9$\pm$1.3 & $\gtrsim$40.1 \\
		& Gradual & PS\#2 & 6.85 -- 7.50 & 7.25 & 
        2.2$\times10^{29}$ -- 1.0$\times10^{27}$ &61&32, 102& $>$121.4 & 
        7.1$\pm$1.2 & 6.5$\pm$0.8 & $>$17.1 \\
		\hline
        \multirow{2}{*}{2023-02-17} & Impulsive & PS\#7 & 6.75 -- 7.30 & 7.00 & 
        3.7$\times10^{28}$ -- 2.6$\times10^{26}$ &N/A&N/A& $>$76.6 &
        2.6$\pm$0.2 & 2.3$\pm$0.9 & $>$29.5 \\
		& Gradual & PS\#3 & 6.90 -- 7.35 & 7.10 & 
        3.8$\times10^{28}$ -- 2.0$\times10^{26}$ &56&45& $>$118.3 & 
        4.3$\pm$0.9 & 4.2$\pm$1.2 & $>$27.5 \\
		\hline\hline
	\end{tabular}
    }}
\end{table}

\begin{figure}[htbp]
	\centering
	\includegraphics[width=\textwidth]{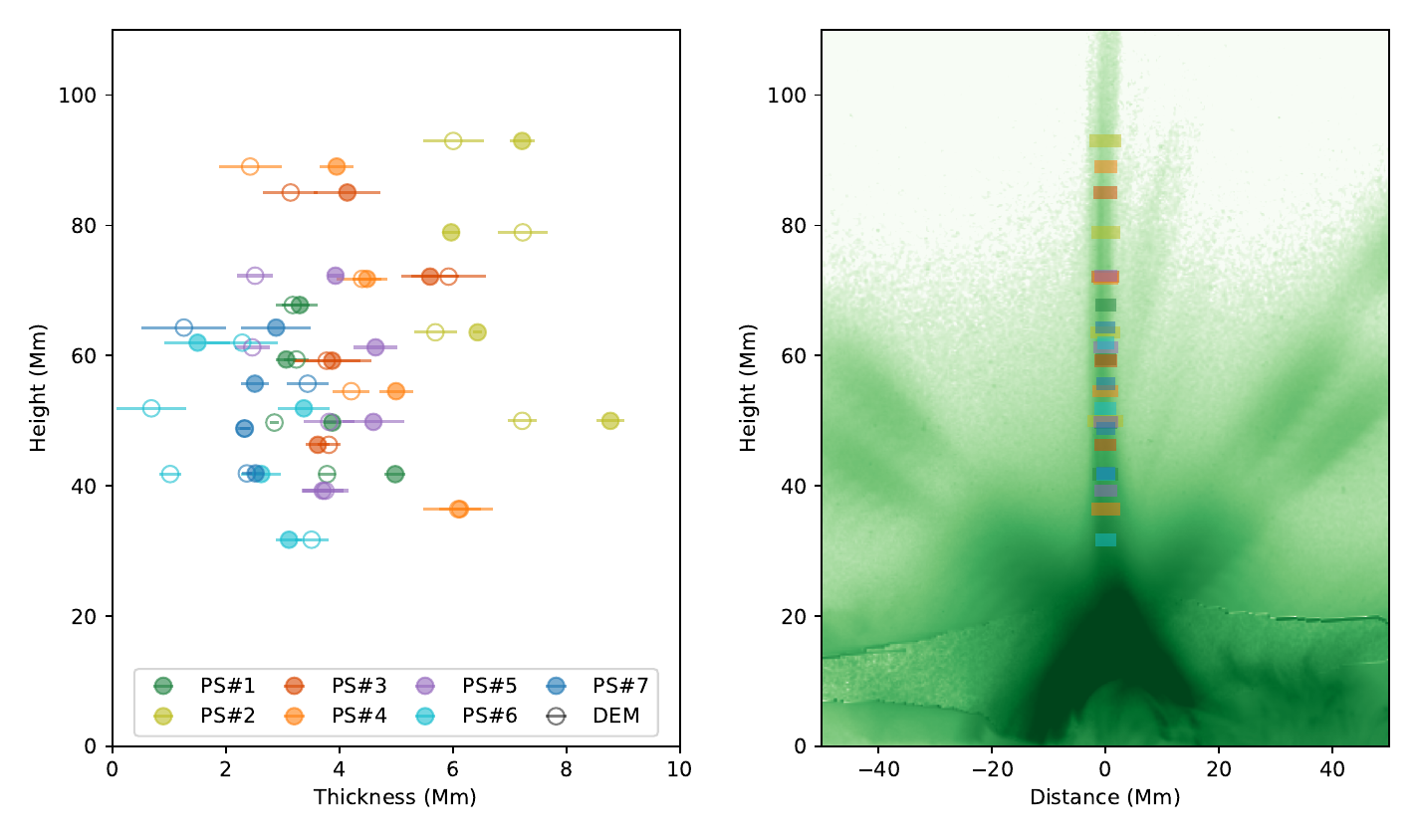}
	\caption{\small Distribution of PS thickness with height. The left panel shows the scatter plot of the thickness as a function of height for the seven PS cases. For each case, the thickness is measured at four different heights from fittings to both the AIA 131~\AA\ intensity (solid circular symbols) and DEM results (hollow circular symbols), which are shown in the same color. The horizontal error bars show uncertainties of the fitting used for the thickness measurement. The right panel shows the thickness distribution with height measured from AIA intensity, overplotted on an AIA 131~\AA\ image of the 2017 September 10 flare (PS\#2) as a visual background. The colors of the thickness bars are the same with those in the left panel.}
	\label{fig:thickness}
\end{figure}

We investigate seven PS features observed during five solar limb flares and analyze their physical properties, including the temperature structure, emission measure, length, and thickness. By isolating cool emissions from the foreground and background, we are able to obtain the accurate temperature and emissions from the PS itself, and measure its parameters based on the pure emissions. We list a summary of the measurements in Table~\ref{tab:summary}. 

The long, thin PS features are basically only prominent in hot AIA channels, and our DEM analyses provide evidence of high-temperature plasma emissions associated with PSs. All the PS structures under study almost contain only a narrow temperature range of hot emissions, suggestive of an isothermal feature for the plasmas surrounding the flare current sheet. Most of the PS features (6 out of 7 cases) exhibit similar properties, which are associated with a high temperature of $\log T_{\rm peak}$=7.0--7.1, a EM range of $10^{26}-10^{28}$~cm$^{-5}$, and a mean thickness of 2--4~Mm. The only different case is PS\#2, the one observed during the gradual phase of the 2017 September 10 flare, which is distinctly hotter ($\log T_{\rm peak}$=7.25), denser (one magnitude higher in \emps), and thicker ($\sim$7~Mm) than the others. We note the 2017 September flare also holds a much higher level of peak SXR flux (GOES-class X8.2) and a much faster CME ($>$3000~km/s) than other cases (Table~\ref{tab:flares}). Another notable feature is that the PS structures exhibit almost constant temperatures at different heights, except for, again, the PS\#2, whose temperature declines slightly with an increasing height above the post-flare loops. 

The flares under study include those having slight different view angles from the edge-on perspective (e.g., PSs \#3, \#4, \#5), which do not show a perfect linear feature like PS\#1 and PS\#2 but exhibit as an interesting 3D structure surrounded by some faint emissions in the bottom (Figure~\ref{fig:flares}). Our results show PS features with similar temperatures in spite of the differences in view angle and line-of-sight depth. Our study also includes two homologous flares from similar locations on the Sun (PSs\#1\&5, both from AR 11748), which hold very similar properties including temperatures, EMs and thicknesses, in the context that they share similar magnetic configurations. Comparing the PSs observed during different phases of flares (PSs in the 2017 and 2023 flares), one can see that the sheet structure during the impulsive phase is generally cooler and thinner than that during the gradual phase of the same flare. We discuss the results in the following sections.

\subsection{Temperature Structure and Plasma Heating}

By performing DEM diagnosis on the flare plasmas and comparing PS with the nearby locations, we find that the PS feature mostly contains hot emissions from a narrow temperature range. These results suggest that the plasmas surrounding the flare current sheet exhibit an isothermal temperature with all plasmas being heated up into above 10~MK. \citet{Warren2018} presented similar isothermal feature for the PS during the 2017 September 10 flare (PS\#2 in our study) by examining the EM loci curves for background-excluded intensities of each AIA channels and EIS spectral lines (their Figure~8). Since the 2017 flare (PS\#2) is the only case that the PS is observable in all six AIA EUV channels but most PSs are only evident in 131~\AA, the EM-loci approach is difficult to apply to the other cases. On the other hand, it is hard in observations to extract the exact PS emissions by excluding the overlapping foreground and background along the line of sight. In our study, by carefully comparing the DEM distribution of PS with its nearby locations (we placed both parallel and perpendicular slits to the PSs to make comparison), we obtain the same results. Our analyses provide evidence of an isothermal temperature for all the PS cases under study.

By examining the DEM distribution of PSs with height, we find that most of the PS structures show almost constant temperatures at different heights above the post-flare loops. This result suggests a long, uniform structure, with a balance of plasma heating and cooling at different heights along the PS. The 2017 September 10 flare (PS\#2) is the exception in that its temperature declines smoothly with an increasing height, which agrees with previous results \citep[e.g.,][]{LiY2018,Warren2018,Longcope2018}. Considering the existence of significant conductive and radiative cooling processes, the isothermal or even increasing temperature in the plasma sheeets suggests additional plasma heating process occurring in the current sheet region, such as retracting magnetic fluxes associated with reconnection downflows \citep{Longcope2018}, global compression from reconnection inflows \citep{Reeves2019}, local heating from supra-arcade downflows within the plasma sheet \citep{Reeves2017,Li2021}, or suppression of conductive cooling due to turbulence \citep{Xie2023}. We also find in some cases the tip of cusp-shaped flare loops underneath contain hotter emissions than the PS (PSs \#1, \#4, \#6, and \#7), which could be contributed from the heating by a pair of slow-mode shocks attached to the reconnection region \citep{Tsuneta1996}. The different temperature distribution can help distinguish the PS feature from the cusp of flare loops, which are often mixed together in imaging observations.

By isolating the PS emissions from foreground and background contributions, we can better characterize the temperature of the PS feature, using either \mtps weighted only by the hot DEM component or $T_{peak}$ of the hot component; while the conventional mean temperature \mt weighted by all DEM components can significantly underestimate the PS's temperature when the cool background emissions dominate (shown in the temperature maps). Since we obtain almost the same temperature range of hot emissions along the whole PS (except PS\#2), it is reasonable to use the peak temperature of this hot component integrated along the whole length as the PS' temperature (in Table~\ref{tab:summary}). All the temperatures we obtained is about $\log T_{peak}$=7.0--7.1, which is very close to that of the Fe XXI emission line and agrees with AIA 131~\AA\ observations. PS\#2 exhibits decreasing temperatures with height, thus $T_{peak}$ of all emissions along it gives a rough estimation that we can use to compare with other cases. The peak temperature of PS\#2 is significantly higher than other cases and is similar to the temperature of Fe XXIV emission line which dominates in AIA 193~\AA channel during a flare.

The PSs observed during the flare impulsive phase are slightly cooler than those during the gradual phase (the 2017 and 2023 flare cases), suggesting that the plasma heating in the current sheet region is more intense after the flare SXR peak, while during the impulsive phase a large portion of magnetic free energy is converted into kinetic energy. The temperature characteristics during different flare phases can be achieved in future detailed studies by examining the temporal evolution of PS temperatures \citep[see, for example,][]{Kittrell2024}.

\subsection{Emission Measure and Plasma Density}

The bright PS feature observed in AIA demonstrates that it is higher in emission and density than the surroundings. We measured that the pure EMs from the hot PSs are in the range of $10^{26}-10^{28}$~cm$^{-5}$ for most cases (4 out of 5 flares), which correspond to electron densities of $10^{8}-10^{9}$~cm$^{-3}$ if assuming a line-of-sight depth of 100~Mm for these X-class flares (taking the 2013 May 13 flare as a reference, which is $\sim$120~Mm). The PS structures during the 2017 September 10 flare show emissions of one magnitude lower (higher) in the impulsive (gradual) phase than other cases, suggestive of varying plasma densities adhering to the reconnection current sheet, particularly unusually dense during the gradual phase. The PSs during the flare impulsive phase hold much less EMs than those during the gradual phase (Table~\ref{tab:summary}; if we consider the superposed hot plasma cloud for PS\#7 in Figure~\ref{fig:ps23i-len}b), which can be resulted from either a shorter line-of-sight depth or a smaller electron density or both. The weak emission of PS during the impulsive eruption makes it more difficult to detect in remote sensing observations.

The emissions of PSs during the flare gradual phase decrease exponentially with height by a factor of two in magnitude. By fitting the profiles using exponential functions, it is clearly seen that the EMs decrease over height much smaller than the scale height under a hydrostatic equilibrium (about 300~Mm for an isothermal temperature of \logt=7.1). EMs of PSs \#3,4,5 generally drop smoothly with height, showing similar trends with the AIA 131 intensities, which can contain significant contributions from the height-dependent background emissions; while the EM--height distributions of PSs \#1,2 are different. We note that PSs \#1\#2 show an almost perfect edge-on view than the cases in other three flares (Figure~\ref{fig:flares}), therefore their \emps provides an better estimation for the hot current sheet plasams.
The hot EMs of PSs \#1\&2 show clearly two-episode evolution with height, where the bottom part drops much steeper than the upper part. In particular, the PS\#2 (2017 September 10 flare) has a scale height for the upper part of 70~Mm larger than the bottom, although the temperature of the former (\logt=6.85--7.2) is significantly lower than the latter (\logt=7.1--7.5). The small scale height in the bottom suggests extremely fast increase in plasma density near the bottom of flare current sheet, which can be caused by downward-moving reconnection outflows that decelerate faster and faster and pile up more toward the bottom, if considering the main reconnection site rises above the visible PS in AIA during the flare gradual phase.

The emissions of PSs during the flare impulsive phase do not show smooth distribution with height but exhibit multiple distinct enhancements, in both AIA 131~\AA\ intensity and EM and even in temperature (Figures~\ref{fig:ps17i-len}, \ref{fig:ps23i-len}). This nonuniformity can be attributed to the existence of multiple plasmoids in the current sheet during fast magnetic reconnection, which is the right case occurring during flare impulsive phase. The plasmoids can have different sizes, experiencing coalescence or further tearing \citep{Shibata2001,Gou2019}. The most notable one in PS\#7 shows a length of about 10~Mm (Figure~\ref{fig:ps23i-len}), which is larger than the PS's thickness and can be the result of a combination of multiple sub-structures. The associated enhancement in temperature (Figure~\ref{fig:ps23i-len}f,g) addresses an interesting aspect for future studies.

\subsection{Sheet Thickness and Magnetic Reconnection}

We measure the thickness of each sheet structure by taking samples at four different heights evenly distributed on the PS. The results are shown in Figure~\ref{fig:thickness}, where the height of each case is calculated with respect to the center of two footpoints of the post-flare loops below. The scatter plot of measured thickness shows no significant height association overall, suggestive of a uniform long sheet structure. The thinnest case is PS\#6 observed during the impulsive phase of the 2017 September 10 flare (cyan symbols; can be thinner than 1~Mm), and the thickest case is PS\#2 observed during the gradual phase of the same flare (yellow; up to $\sim$9~Mm). The thicknesses measured in the impulsive phase of the 2023 flare (PS\#7) are smaller than those in the gradual phase (PS\#3) as well. This result suggests a higher reconnection rate during flare impulsive phase when the current sheet is thinning into smaller scales which may result in multiple reconnection sites. For each individual case, the thickness measured in DEM (hollow symbols) is mostly smaller than that from AIA intensity (solid symbols), where the former gives a better constraint for the hot plasmas.

The thickness is generally consistent with previous results measured in EUV and X-ray observations \citep[e.g.,][]{Savage2010,Seaton2017,LiY2018}. We note it is the apparent thickness of the hot plasmas surrounding the flare current sheet but not the electric current sheet itself, while the latter is expected to be even thinner. The length-to-thickness ratios of PS features (Table~\ref{tab:summary}) are much larger than the threshold of tearing mode instability \citep{Furth1963}, although the length in the study is an underestimation containing only the bottom, visible part of PSs in AIA (except for PS\#6 which shows a full length).

\section{Summary}

We study the thermal properties of PS structures observed in AIA EUV channels, which form in the wake of flare eruptions connecting the post-flare loops and serve as the primary site for magnetic reconnection. The PS features contain only a narrow temperature range of hot plasmas at around \logt =7.0-7.1, and the temperature remains constant at different heights along the PS. The PSs observed during the flare impulsive phase are generally cooler, thinner, and less dense than those during the gradual phase, indicative of different thermal properties in the context of high reconnection rate during the impulsive eruption. Our results show a long, uniform structure with an isothermal temperature, suggesting balanced heating and cooling processes along the sheet, particularly associated with additional plasma heating. The 2017 September 10 flare of the highest GOES-class shows an exceptional distribution that the PS's temperature increases toward the flare looptop during the gradual phase; while it is hotter, denser, thicker than other events, and it also extends high beyond the AIA's FOV. When a magnetic flux rope erupts into the high corona, the EUV sheet structure observed in AIA is only the bottom portion of a long current sheet connecting to the runaway eruption, thus it is also worthy of investigating how the properties of the complete structure are distributed. Such studies can benefit from observations of the extended corona for example by GOES/SUVI and future missions such as ECCCO. Although it is well-accepted that the reconnection current sheet is highly dynamic and fragmented with numerous different-scale substructures, its surrounding plasmas behave as a steady and uniform structure in terms of thermal properties in the macroscopic scale. To complement the thermal characteristics, spectroscopic diagnostics from hot coronal emission lines are particularly advantageous to understand the nonthermal and turbulent processes in the flare current sheet.

\begin{acknowledgments}
We are grateful to the NASA SDO/AIA science team for the science data and analysis tools. TG acknowledges the support by contract SP02H1701R from Lockheed-Martin to SAO. KKR acknowledges support from NASA grant 80NSSC19K0853.
\end{acknowledgments}

\bibliography{cs}{}
\bibliographystyle{aasjournal}

\end{document}